\documentclass[twocolumn,prl,aps,floatfix,superscriptaddress]{revtex4-1}
\usepackage{bm,color,amsmath,amssymb,mathrsfs,latexsym,graphicx,float}

\usepackage{psfrag}	

\usepackage[caption=false]{subfig}
\usepackage{times}
\usepackage{amsmath}
\usepackage{epstopdf}
\usepackage{amsfonts}

\usepackage{graphics}
\usepackage{ulem}

\newcommand{\be}{\begin{equation}}
\newcommand{\ee}{\end{equation}}
\newcommand{\bea}{\begin{eqnarray}}
\newcommand{\eea}{\end{eqnarray}}

\renewcommand{\phi}{\varphi}
\renewcommand{\epsilon}{\varepsilon}

\begin{document}
\title{Boundary states of 4D topological matter: Emergence and full 3D-imaging of nodal Seifert surfaces}

\author{Linhu Li}  \email{phylli@nus.edu.sg}
\affiliation{Department of Physics, National University of Singapore, Singapore 117551, Republic of Singapore}\author{Ching Hua Lee} \email{calvin-lee@ihpc.a-star.edu.sg}
\affiliation{Institute of High Performance Computing, A*STAR, Singapore, 138632.}
\affiliation{Department of Physics, National University of Singapore, Singapore 117551, Republic of Singapore}
\author{Jiangbin Gong}  \email{phygj@nus.edu.sg}
\affiliation{Department of Physics, National University of Singapore, Singapore 117551, Republic of Singapore}

\date{\today}

\begin{abstract}
With its boundary tracing out a link or knot in 3D, the Seifert surface is a 2D surface of core importance to topological classification. We propose the first-ever experimentally realistic setup where Seifert surfaces emerge as the boundary states of 4D topological matter. Unlike ordinary real space knots that exist in polymers, biomolecules and everyday life, our knots and their Seifert surfaces exist as momentum space nodal structures, where topological linkages have profound effects on optical and transport phenomena.
Realized with 4D circuit lattices, our nodal Seifert systems are freed from symmetry constraints and readily tunable due to the dimension and distance agnostic nature of circuit connections. Importantly, their Seifert surfaces manifest as very pronounced impedance peaks in their 3D-imaging via impedance measurements, and are directly related to knot invariants like the Alexander polynomial and knot Signature. This work thus unleashes the great potential of Seifert surfaces as sophisticated yet accessible mathematical tools in the study of exotic band structures.
\end{abstract}

\maketitle
\section{introduction}
The irresistible allure of topological physics has brought together generations of physicists and engineers in witnessing how abstract beauty and experimental pragmatism coincide. In higher dimensions especially, the language of topology enables the understanding of novel and unexpected phenomena in terms of universal and robust motifs. A quintessential example is given by nodal knots existing in momentum space, where the knotted structure leads to new phases of matter protected by topological knot invariants.
Unlike knotted molecules or optical vortices in real space\cite{bio,bio2,Seifert_light}, nodal knots consist of valence and conduction bands intersecting along one-dimensional (1D) lines in momentum space, which intertwine to form
knotted nodal loops (NLs)
~\cite{linkknot1,linkknot2,linkknot3,linkknot4,linkknot5,linkknot6,linkknot7,Li2018,lee2019imaging,stalhammar2019hyperbolic} so multifarious that topological invariants take the form of polynomials rather than the $\mathbb{Z}_2$ or $\mathbb{Z}$ integers~\cite{Li2018,lee2019imaging,stalhammar2019hyperbolic,KnotTheory,Collins_Seifert} of ordinary topological insulators. Fundamental in constructing such invariants are the Seifert surfaces bounded by the nodal structure~\cite{KnotTheory,Collins_Seifert}, which assume interesting, bubble-like shapes demarcating ``drumhead'' topological regions in the projected 2D surface Brillouin zone (BZ)~\cite{Burkov2011,Weng2015,Kim2015,Yu2015}.

As compact and orientable surfaces bounded by nodal knots or links, Seifert surfaces not only provide convenient visualization, but are also of core importance to topological classification. The linking properties of their homology generators can be used to compute~\cite{Li2018,Collins_Seifert} the Alexander polynomial - a classical knot invariant - of the NL or knot, hence distinguishing it from other nodal configurations.
Indeed, Seifert surfaces are central to knot theory and low-dimensional topology~\cite{LDT}, provoking many fascinating mathematical and computational problems, such as the uniqueness of a minimal genus Seifert surface~\cite{Seifert1}, and their construction and visualization~\cite{Seifert2,Seifert3}.
Its geometric appeal, e.g., appearing as a twisted band for the Hopf link, has also engendered much interest in other subfields, with alternative interpretations as contours of constant \textit{real space} optical polarization azimuths~\cite{Seifert_light} and dissipationless ``Fermi" surfaces~\cite{nonHermitian_NL1,nonHermitian_NL2,nonHermitian_NL3}.

 Despite their mathematical exuberance, Seifert surfaces do not naturally emerge from static 3D systems. To date, only their shadows (drumhead states)
 on the 2D surface BZ of a 3D topological matter has been connected with physical measurements.
 To overcome this and to unleash the full potential of Seifert surfaces in advancing the classification of topological matter, we lend our inspiration from recent advances in synthetic higher dimensional topological matter, e.g. 4D quantum Hall systems 
\cite{HD_exp1,HD_exp2,HD_exp4,HD_exp5}, where the additional dimensions bring theoretical novelties~\cite{higher_dimension1,higher_dimension2} like 5D Weyl semimetals~\cite{higher_dimension3,higher_dimension4} close to physical reality. Specifically, we shall design 3D NLs embedded in \textit{parent} 4D nodal structures, such that Seifert surfaces naturally emerge as topologically robust zero-energy surfaces at their 3D boundaries. In
essence, we propose to embed 3D NLs or their resultant knots
in a 4D setup such that all desired NL structures are respectively
associated with different quasimomentum values along
the 4th dimension. Upon open boundary condition (OBC) taken along the 4th dimension,
all such NL structures collapse onto the same 3D BZ and hence more complicated NL linkage or knots can be created.

{Having the 4th dimension makes the momentum space nodal topology much more experimentally accessible through Seifert surface imaging, even in the face of added complexity.
Unlike their 3D counterparts, 4D NL systems do not require any sublattice symmetry, and the 2D  Seifert surfaces can be reconstructed more easily, compared with 1D NLs as thin structures detectable only at extremely high momentum-space resolution.
More interestingly, arbitrarily many NLs can be systematically encapsulated in the 3D ``boundary" Brillouin zone of a \textit{single} 4D system with relatively simple coupling configurations.
As we will demonstrate, such 4D systems are most suitably implemented via RLC circuit setups, where lattice sites are simulated by the circuits nodes, and positive/negative couplings between them by capacitors/inductors.
Compared to existing higher dimensional optical systems with synthetic dimensions, circuit implementations have the advantages of being extremely versatile, inexpensive and reconfigurable~\cite{circuit1,circuit2,circuit3,circuit4,circuit_Luo,circuit_Hadad,circuit_Zhu,circuit_Goren,helbig2018band,hofmann2018chiral,wang2018topologically}, with nodes connected in any desired way free from constraints of locality or dimensionality.
This versatility in implementing a ``genuine" physical 4th dimension is crucial in obtaining our topological boundary Seifert surfaces, which cannot exist in approaches where time takes the role of the 4th dimension~\cite{NL_4D}.

}

\section{results}
\subsection{Drumhead states versus Seifert surfaces}
We begin by clarifying the exact relationship between the 2D ``drumhead'' surface states of 3D nodal systems, and the 2D Seifert surface states within the 3D boundary of a 4D nodal system. Consider a minimal 2-band ansatz Hamiltonian
\begin{eqnarray}
h({\bm k})=h_0({\bm k})\,\mathbb{I}+\sum_{i=1}^3 h_i({\bm k})\sigma_i,\label{minimal}
\end{eqnarray}
with $\sigma_i$ the $i$-th Pauli matrix acting in a pseudospin-1/2 space, ${\bm k}$ being the quasi-momentum vector.  Nodes (pseudospin singularities) occur when $h_i(\bm k)=0$ for all $i=1,2,3$, such that the conduction and valence bands touch. In 3D, the nodes form NLs only when one of $\sigma_i$ is constrained to be zero, typically by sublattice symmetry (e.g. $\sigma_3=0$). But in 4D, NLs occur generically without \textit{any} symmetry requirement, since the three constraints $h_i(\bm k)=0$ still leave a nodal solution set with codimension 1.
At first glance, these 4D NLs do not seem interesting since nontrivial knots and links only exist in 3D. Yet, as we shall shortly show, the 3D boundary of such 4D nodal systems exhibits spectacular promise for the practical imaging of nodal knots.



\begin{figure}
\includegraphics[width=1\linewidth]{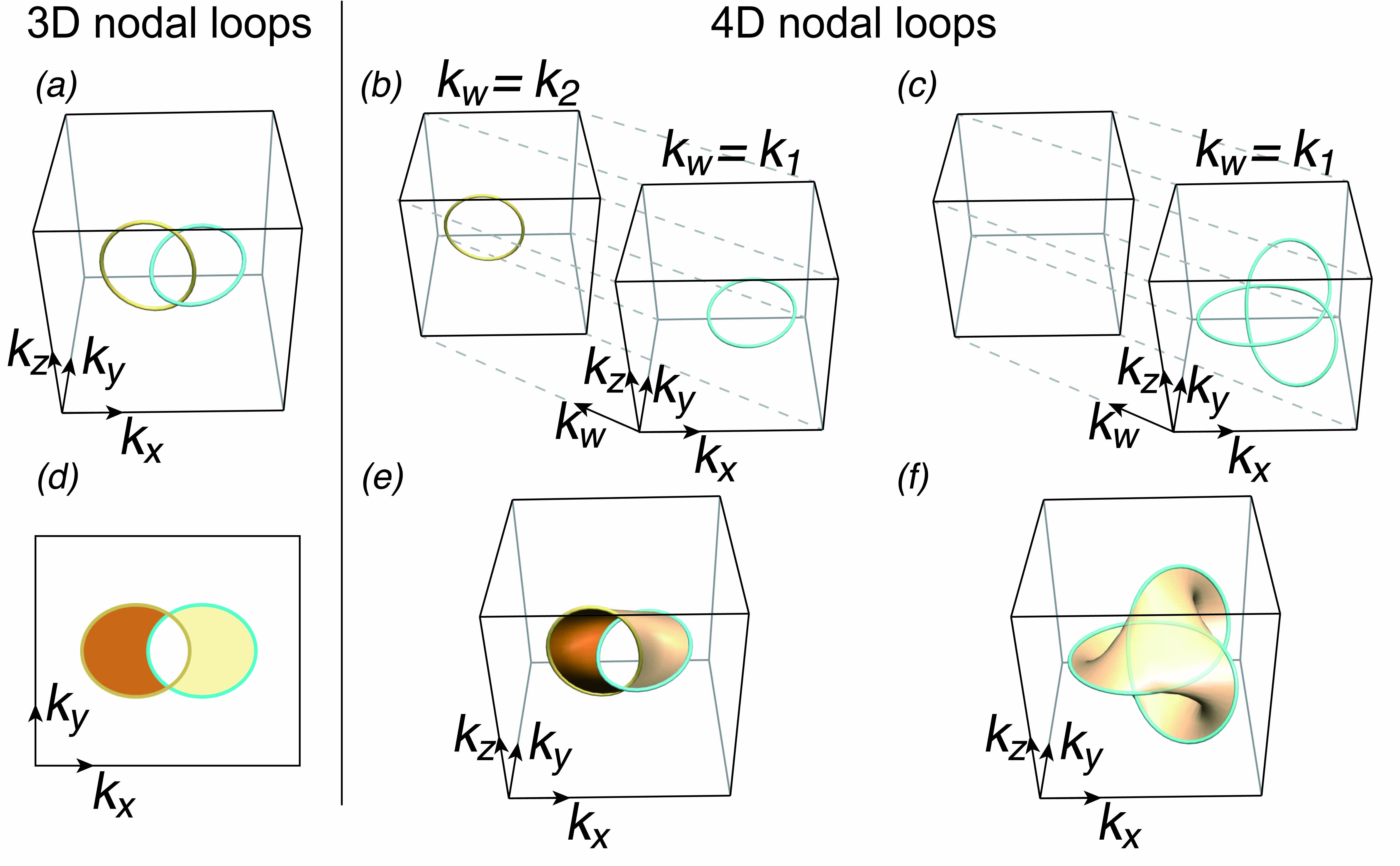}
\caption{The relation between 3D (a) and 4D (b-c) NLs and their respective boundary drumhead states (d) and Seifert surfaces (e-f). Panels (d-f) depict the boundary BZ under OBCs, with yellow and blue curves corresponding to projected NLs and brown shaded regions corresponding to topological boundary states. 
}
\label{fig1}
\end{figure}

{Consider first the drumhead states in 3D nodal systems}. Under OBCs, a typical 3D nodal system exhibits drumhead surface (2D boundary) states that fill the 2D region enclosed by the surface-projected NLs/knots [Fig.\ref{fig1}(a,d)], with dispersion given by $h_0(\bm k)$. Essentially, drumhead states are boundary projections of a bulk surface stretched across the NLs i.e. a taut Seifert surface~\cite{KnotTheory,Collins_Seifert} of the NLs, with degeneracy corresponding to the multiplicity of the projection. But it has to be emphasized that this Seifert surface of a 3D nodal system is an entirely abstract construction~\cite{footnote1} not verifiable from 2D boundary states, insofar as 3D geometric information, particularly of the knot over/under-crossings, is already lost in the surface projection.


For \textit{physically} realizing Seifert surfaces as boundary states and hence directly observing the knot topology, we consider 4D nodal systems defined in the 3 ordinary dimensions plus an additional dimension labeled by $w$. The key inspiration is that although the NLs are always unlinked and unknotted in 4D, they can be linked or knotted when ``compressed'' into 3D via (3D) boundary projection. This being the case, the topological boundary states of the given 4D topological matter, which interpolate the interior of the NLs, must necessarily form a Seifert surface embedded in a physical 3D BZ and terminating at NLs. This is illustrated in Fig.~\ref{fig1}(b,c) and (e,f) with OBCs along the $\hat{w}$ direction. In Fig.~\ref{fig1}(b) for instance, the 4D nodal structure is chosen to consist of two unlinked NLs embedded in their respective 3D BZ subspaces indexed by $k_w=k_{1,2}$, the quasi-momenta labeling their slice in the 4th dimension.  With $k_w$ projected out by OBCs, the two NLs become nontrivially linked in the 3D boundary BZ, and are interpolated by a Seifert Fermi (zero energy) surface [Fig.~\ref{fig1}(e)].
Similarly, a nodal Trefoil, which is unknotted in 4D [Fig.~\ref{fig1}(c)], becomes knotted when projected into a 3D boundary [Fig.~\ref{fig1}(f)].
Alternatively, one may understand such 4D NL systems as 3D Weyl systems equipped with an additional dimension, such that Weyl points and their Fermi arcs trace out NLs and Seifert surfaces respectively along the additional dimension. With this insight, we can associate some exotic behaviors of Fermi arcs with the nontrivial topology of their parent NLs projected onto the 3D surface Brillouin zone, as discussed later when explicit constructions of nontrivial links and knots are introduced.

\begin{figure}
\includegraphics[width=1\linewidth]{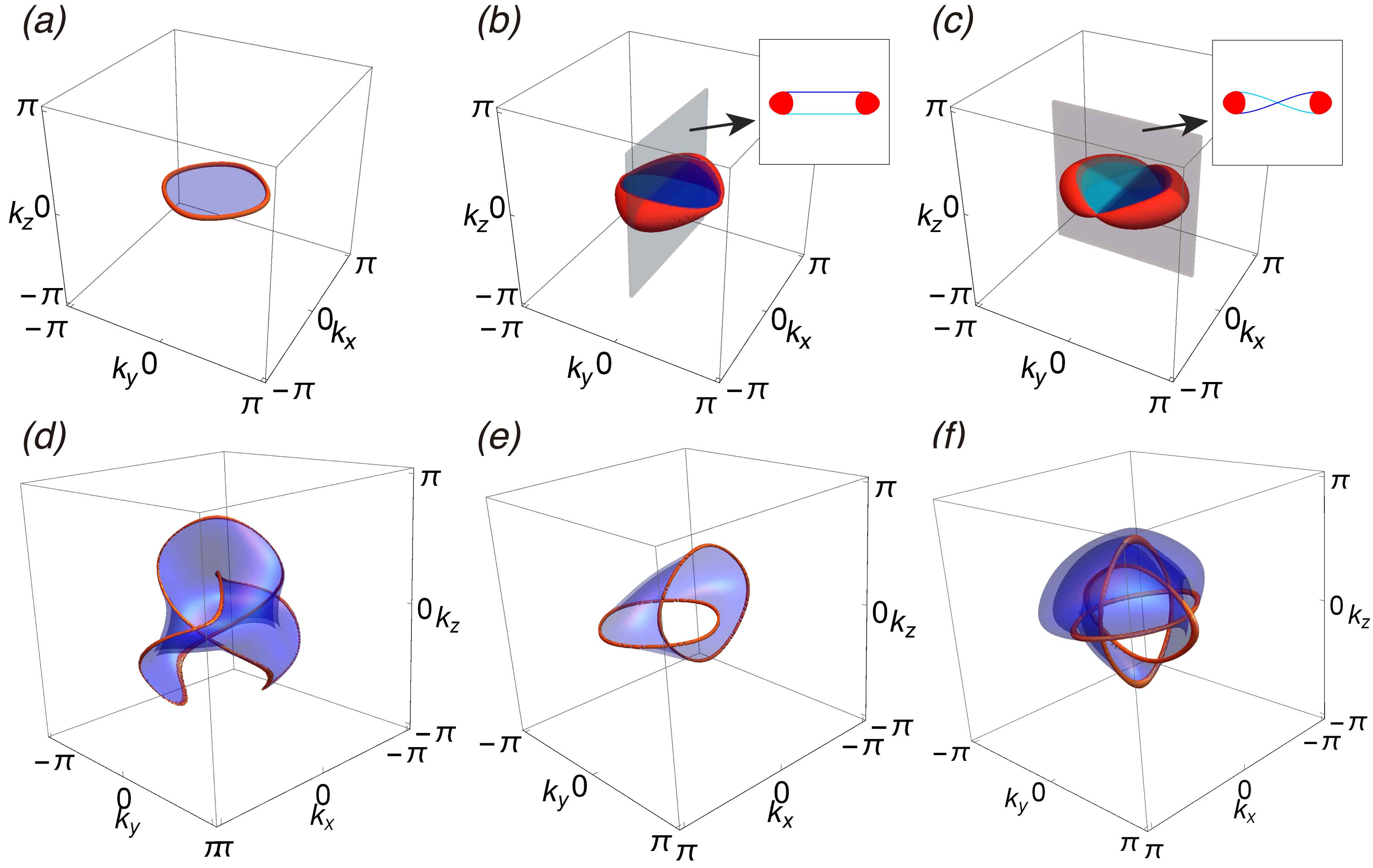}
\caption{
(a-c) Boundary states of a single 4D NL (Eq.~\ref{1NL}) with $h_0({\bf k})=0$ in (a), $h_0({\bf k})=0.4\cos k_y$ in (b), and $h_0({\bf k})=0.4\sin k_y$ in (c), demonstrating the robustness of the Seifert surface against $h_0$ perturbations. Red regions represent zero-energy bulk states (nodal solutions) whereas dark and light blue regions depict boundary states as Seifert surfaces of the NLs.  Note that in panel (c), the light blue and dark blue regions are partially covered up by each other.
(d) Boundary Trefoil knot and its Seifert surface states from Eq.~\ref{h4D}. (e-f) Boundary Hopf-link and Borromean rings given by Eq.~\ref{h4DN} and their Serfiet surfaces, with $N=2$ and $N=3$ linked loops respectively.
}\label{fig2}
\end{figure}

Before discussing general routes to topologically nontrivial Seifert surfaces as boundary states, we first explicitly describe the simplest possible 4D Hamiltonian possessing a single NL: 
\begin{eqnarray}
h_{1{\rm NL}}({\bf k})&=&(\cos{k_x}+\cos{k_y}+\cos{k_z}+\cos{k_w}-m)\sigma_a\nonumber\\
&&+(\sin{k_w})\sigma_b+(\sin{k_z})\sigma_c + h_0({\bf k})\,\mathbb{I},
\label{1NL}
\end{eqnarray}
with $x,y,z,w$ labelling the 4 dimensions,  and $\sigma_a,\sigma_b,\sigma_c$ an arbitrary permutation of the three Pauli matrices. 
While the nodal structure is agnostic to the Pauli matrix basis, practical implementations may require specific choices dictated by symmetry. In this work, we assume no specific basis except when discussing the circuit realizations, where time-reversal symmetry holds. 
When $2<m<4$, Eq.~\ref{1NL} describes a single NL $\cos k_x +\cos k_y =m-2 $ in the $k_w=k_z=0$ plane [Fig. \ref{fig2}(a)].
Under $\hat w$-direction OBCs, topological boundary states must appear due to the bulk-edge correspondence
associated with a nontrivial Chern number, as shown in the Supplementary Materials \cite{SuppMat}. Those boundary states at zero energy
then make up the Seifert surface $\cos k_x +\cos k_y <m-2 $, $k_z=0$ matching the identified NL (blue).

Compared to the 1D NLs in the 4D BZ, Seifert surfaces are experimentally more robust for various reasons. Firstly, they are 2D surfaces in the 3D surface BZ, and are thus easy to image even at low resolutions. Secondly, they are localized at the 3D boundaries of a 4D lattice, which are relatively accessible by external probes. Thirdly, freed from the requirement of sublattice symmetry, they behave as chiral boundary states of 2D QH systems with two other momenta as system parameters [See Fig. S1 in the Supplementary Materials \cite{SuppMat}], and are thus immune to 
extra terms induced by noise or spatial modulations. Consider for instance a perturbation in $h_0({\bf k})$.  In 3D nodal systems, such terms will introduce momentum dependence in the energy and destroy the flatness of drumhead states and hence the boundary Fermi surface. However, in 4D nodal systems, they merely deform the boundary zero-energy surface in momentum space, which persist as 2D (zero-energy) Seifert surfaces of the NLs, thus being robust to the perturbation.  Shown in Fig. \ref{fig2}(b) and (c) are two illustrative examples: $h_0({\bf k})\propto\cos k_y$ and $h_0({\bf k})\propto\sin k_y$. In the former case, the Fermi surfaces belonging to the two opposite OBC boundaries are 
fully separated and displaced in opposite directions, while in the latter they intersect along a line.

\subsection{Nodal knot Seifert surfaces}
We first show how a \textit{single} nodal knot and its Seifert surface can be generically realized in the 3D boundaries of 4D NL systems. Starting from an ordinary NL system Hamiltonian defined in 3D,\begin{eqnarray}
h_{\rm 3D}({\bf k}_{\rm 3D})=h_a({\bf k}_{\rm 3D})\sigma_a+h_b({\bf k}_{\rm 3D})\sigma_b,\label{nodal_knot_H}
\end{eqnarray}
 we can always construct a 4D NL system Hamiltonian
\begin{eqnarray}
h_{\rm 4D}({\bf k})&=&h_{\rm 3D}({\bf k}_{\rm 3D})+h_w(k_w),\label{h4D}\\
h_w(k_w)&=&t_w\left[(\cos{k_w}-1)\sigma_a+(\sin{k_w})\sigma_c\right],
\label{h4Dw}
\end{eqnarray}
with ${\bf k}=({\bf k}_{\rm 3D},k_w)=(k_x,k_y,k_z,k_w)$ and $t_w$ setting the scale of $h_w$. We shall offer two perspectives for understanding this resultant 4D system. From the first perspective, it may be understood as a 1D superlattice with only nearest neighbor couplings along the $\hat w$ direction, such that each supercell is a 3D system $h_{\rm 3D}({\bf k}_{\rm 3D})$ that contains the desired nodal knot (whose Seifert surface is yet to be revealed).
Interestingly, $h_{\rm 3D}$ and $h_{\rm 4D}$ can be made to contain exactly the same bulk NLs. To see this, note that the gap of $h_{\rm 4D}({\bf k})$ closes when 
\begin{eqnarray}
k_w&=&0,~~h_a^2({\bf k_{\rm 3D}})+h^2_b({\bf k_{\rm 3D}})=0,~{\rm or}\label{con1}\\
k_w&=&\pi,~~[h_a({\bf k_{\rm 3D}})-2t_w]^2+h^2_b({\bf k_{\rm 3D}})=0.\label{con2}
\end{eqnarray}
By choosing $2t_w>{\rm max}[h_a({\bf k_{\rm 3D}})]$, Eq.~(\ref{con2}) is never satisfied, and so the gap closure conditions [Eqs.~(\ref{con1}) and (\ref{con2})] for $h_{\rm 4D}({\bf k})$ reduce to that of $h_{\rm 3D}({\bf k}_{\rm 3D})$.

To gain more insights, we shall introduce the second perspective, where we divide the four dimensions into two groups, namely, $(\hat{z}, \hat{w})$ and $(\hat{x}, \hat{y})$, such that $h_{\rm 4D}({\bf k})$ can be viewed as a 2D system in $(\hat{z}, \hat{w})$ dimensions with
$k_x$ and $k_y$ serving as two system parameters. Now if the bulk  Hamiltonian
$\left[h_a({\bf k}_{\rm 3D})+t_w (\cos{k_w}-1)\right]\sigma_a+t_w(\sin{k_w})\sigma_c+h_b({\bf k}_{\rm 3D})\sigma_b$ features a nonzero topological Chern number, the chiral boundary states must emerge upon taking OBCs along the $\hat w$ direction, with the emergence of
zero-energy boundary states requiring the obvious chiral-symmetry condition $h_b(\bf k_{\rm 3D})=0$. Moreover, the second condition $h_a({\bf k}_{\rm 3D})=0$ represents the topological phase transition condition for such boundary states to appear (see the Supplementary Materials \cite{SuppMat}). All such zero-energy boundary states parameterized by $(k_x, k_y)$ form a bona-fide Seifert surface (albeit not necessarily the minimal area Seifert surface)
matching the nodal knot
as the intersection of $h_a(\bf k_{\rm 3D})=0$ and $h_b(\bf k_{\rm 3D})=0$ surfaces.
Illustrated in Fig. \ref{fig2}(d) is the boundary Seifert surface of a Trefoil knot, with its $h_{\rm 3D}({\bf k}_{\rm 3D})$ and the designed $h_{\rm 4D}({\bf k})$ detailed in Supplementary Materials. The key takeway of this construction is that, by connecting identical copies of 3D NL supercells with nearest neighbor couplings, one can realize not just the same NLs, which are elusive to image, but also their Seifert surfaces which are easier to image, being extensive in one additional dimension. Such nearest neighbor couplings are easy to implement with circuits, as discussed later.


\subsection{Seifert surfaces of arbitrarily many linked NL components }

4D extension can furthermore link \textit{arbitrarily many} of such nodal structure components and their Seifert surfaces without increasing real-space complexity. Like illustrated in Fig. 1(b), the 4th dimension allows multiple 3D NLs in different $k_w$ subspaces to be embedded in the same 4D NL system. 
Given $N$ different 3D NLs possessed by $h_{n,\rm 3D}({\bf k}_{\rm 3D})=h_{n,a}({\bf k}_{\rm 3D})\sigma_a+h_{n,b}({\bf k}_{\rm 3D})\sigma_b$, $n=1,...,N$, a 4D NL system that encapsulates them all can be constructed as follows:
\begin{eqnarray}
h_{N,\rm 4D}({\bf k})=\sum_{n=1}^{N}h_{n,\rm 3D}({\bf k}_{\rm 3D}) g_n(k_w)+f(k_w)\sigma_c.
\label{h4DN}
\end{eqnarray}
Here $f(k_w)=0$ at $N$ values of $k_w$, i.e., $k_w=k_{w,n}$, with $n=1,2,...,N$.  Provided that each $g_{m}(k_{w})$ at $k_w=k_{w,n}$
is nonzero when and only when $n=m$, the band touching condition for $h_{N,\rm 4D}({\bf k})$ then yields a collection of all the $N$ NLs we start with.
Under OBCs in the $\hat w$-direction, all these NLs collapse into the same 3D boundary BZ, forming an intricately linked structure with $N$ nodal components. As described in Supplementary Materials with minimal choices for $f(k_w)$ and $g_n(k_w)$, the topological boundary states of $h_{N,\rm 4D}$ consist of Seifert surfaces of $N$ linked nodal structures.


Illustrated in Fig. \ref{fig2}(e) and (f) are two examples of Seifert surfaces with $N=2$ and $N=3$, corresponding to a Hopf-link and a set of Borromean rings respectively, with detailed Hamiltonians given in the Supplementary Materials.
The latter NL system has the curious property that each pair of loops is unlinked, even though the nodal structure has a nontrivial linkage characterized by the Milnor number~\cite{KnotTheory}. Despite their intricacy, multiply linked NLs like such can always be realized as a superlattice with only \textit{nearest neighbor} couplings along the 4th dimension. The complexity is relegated to the interior structure within each supercell, which implements $f(k_w)$ and $g_n(k_w)$ in a modular and hence experimentally convenient manner, as discussed in the following section of circuit realization.

\subsection{Relation of Seifert surfaces to Fermi arcs}
As discussed in the previous section and the Supplementary Materials, the nodal loops and the Seifert surface states originate from 2D Chern topology and exist without symmetry restrictions. Thus they provide 4D analogs of the 3D Weyl semimetals and Fermi arcs, but with richer topological structures of knots and links. 4D NL systems can also be conceptualized as a 3D Weyl systems extended along an additional {fourth} dimension, such that Weyl points and their Fermi arcs trace out NLs and Seifert surfaces respectively {along the fourth dimension}. Therefore, we can associate some exotic behaviors of Fermi arcs with the nontrivial topology of their parent NLs projected onto the 3D surface Brillouin zone.

\begin{figure}
\includegraphics[width=1\linewidth]{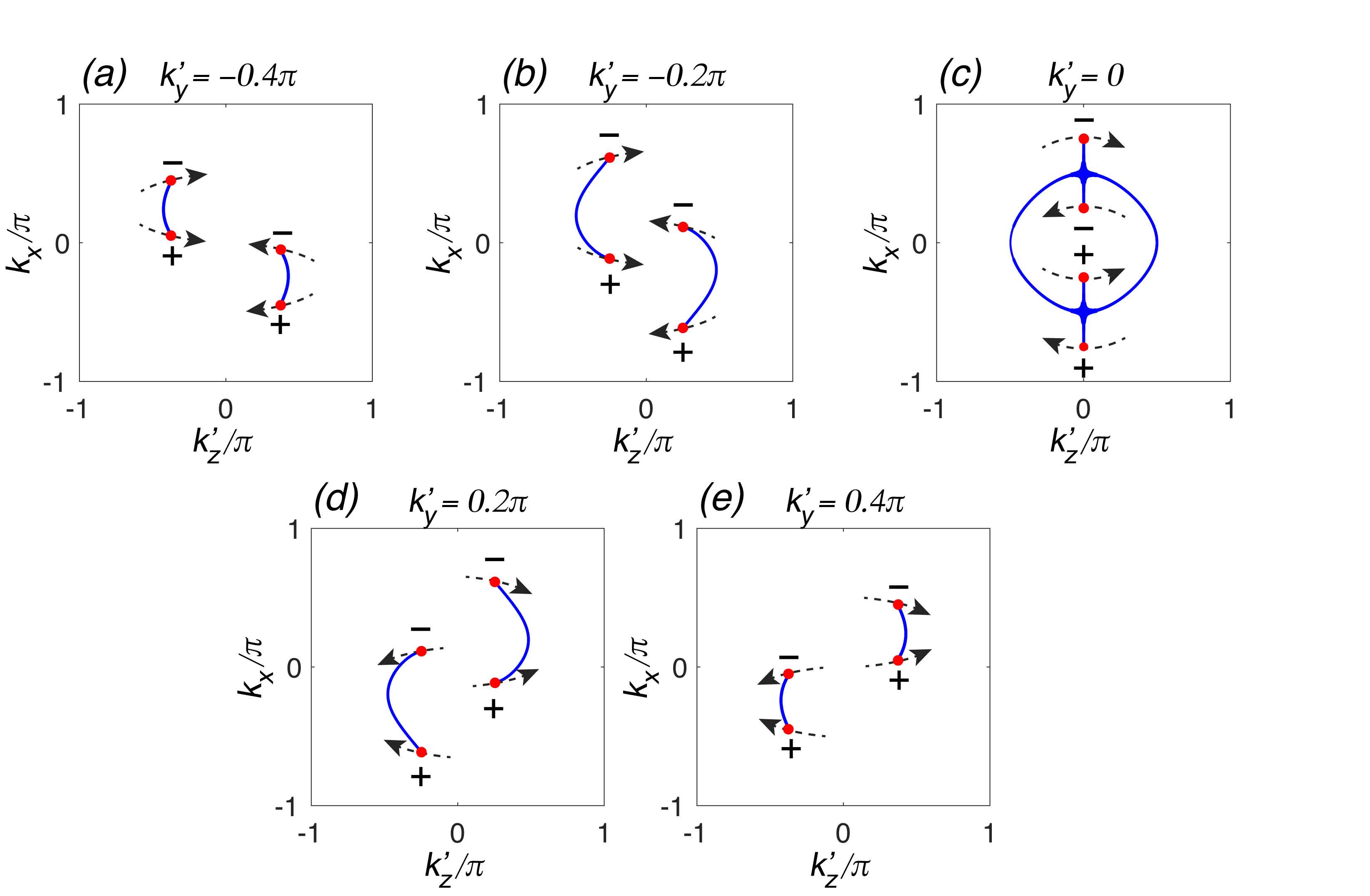}
\caption{
Weyl points (red points) and Fermi arcs (blue lines) of a 3D system given by the model of
Fig. \ref{fig2}(e) with $k'_y=k_y+k_z$, $k'_z=k_y-k_z$, and $k'_y$ taken as a parameter describing the additional {fourth} dimension.
The parameters are $m=2$ and $\phi=\pi/4$. The arrows show the movement of the Weyl points when increasing $k'_y$. The chirality of the Weyl points is shown by the plus or minus signs in the figure.
}\label{fig:Fermi_arc}
\end{figure}

As illustrated in Fig. \ref{fig:Fermi_arc}, 
the 4D Hopf-link Hamlitonian of Fig. \ref{fig2}(e) with $k'_y=k_y+k_z$, $k'_z=k_y-k_z$ has $k'_y$ taken as a parameter describing the additional {fourth} dimension. With OBC along $\hat{w}$, the two Fermi arcs connecting different pairs of Weyl points can move in the 2D plane of $k_x$ and $k'_z$, and exchange portions of their arcs when $k'_y$ varies through zero.
Specifically, the Fermi arcs touch each other and form an exotic crossed flatband \cite{BoLu2015crossed} at $k'_y=0$ [Fig. \ref{fig:Fermi_arc}(c)].

\subsection{Imaging Seifert surfaces through circuit impedance measurements}
Having described the mathematical construction of nodal Seifert surfaces, it is hence important to find an experimentally feasible realization of our approach.
Below we  discuss how {a nontrivial link} can be robustly realized and measured in an electrical circuit setup. Circuit realizations enjoy several advantages: 1) circuit connections are incredibly versatile, with coupling networks of arbitrarily non-locality or high dimensionality easily realizable with suitable wire configurations, 2) 3D boundary terminations are easily accessible as surface nodes of a circuit network and, perhaps most importantly, 3) massive Seifert Fermi surface degeneracies are easily detectable as pronounced ``topolectrical'' resonances already observed in other contexts~\cite{circuit1,circuit2,circuit3,circuit4,circuit_Luo,circuit_Hadad,circuit_Zhu,circuit_Goren,helbig2018band,hofmann2018chiral,wang2018topologically}.

\begin{figure*}
\includegraphics[width=0.9\linewidth]{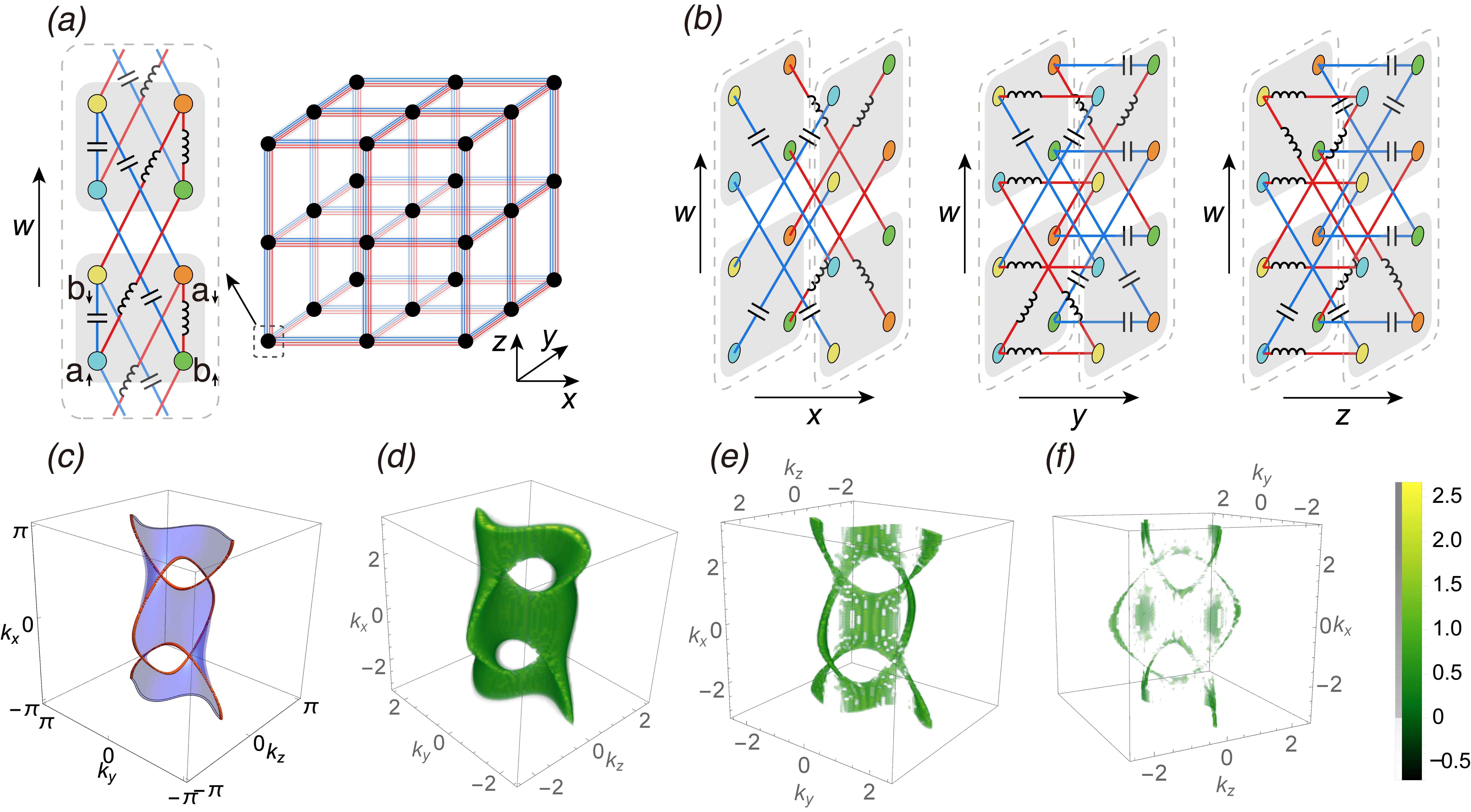}
\caption{a) Sketch of our 2-link circuit (Eq.~\ref{circuit_link2}) in terms of supercell internal structure (Left) and overall connectivity (Right). Gray shadows demarcate individual unit cells within a supercell, with each of the four sublattices colored differently, and red/blue lines indicating positive/negative couplings implemented by inductors/capacitors, as detailed in Supplementary Materials. (b) Detailed illustrations of inter-supercell circuit couplings along the $x$, $y$, and $z$ directions. (c) Analytically computed NLs (red loops) that bound a topologically robust Seifert surface (blue region) of the 2-link with $m=1.5$, which is accurately reconstructed from topolectrical resonance simulations via Eq.~\ref{IJV2}. Impedances across intra-unit cell diagonal sublattices [(d) $\log{|Z_{11}({\bm k}_{\rm 3D})|}$, (e) $\log{|Z_{22}({\bm k}_{\rm 3D})|}$, and (f) $\log{|Z_{33}({\bm k}_{\rm 3D})|}$] were computed with realistic $1\%$ disorder. From the 3D surface (d) towards its 4D bulk (f), the Seifert surface gradually decays into the bulk NL.}
\label{lattice_circuit1}
\end{figure*}

Unlike a quantum mechanical lattice governed by Schr\"{o}dinger's equation, a circuit network is governed by Kirchhoff's equation. In a matrix form, Kirchhoff's law yields $I_\mu=\sum_\nu J_{\mu\nu}V_\nu$, where $I_\mu$, $V_\mu$ are vectors with components representing the input current and electrical potential at node $\mu$.  The circuit Laplacian $J_{\mu\nu}$, which expresses the input currents in terms of the potentials, replaces the role of the Hamiltonian in determining the spectrum relevant to the impedance. In a standard RLC circuit at AC frequency $\omega$, the resistors, inductors and capacitors respectively contribute off-diagonal terms $-R^{-1}$, $-(i\omega L)^{-1}$ and $-i\omega C$ to the Laplacian~\cite{circuit3}, consistent with the time-reversal symmetry condition $J^*({\bm k})=-J(-{\bm k})$. Since this mandates that any NLs must be symmetric in $\pm{\bm k}$, we shall frequently realize NLs in mirror-image pairs, such as those detailed in the Supplementary Materials \cite{SuppMat} for circuit realizations of unlinked nodal rings and a pair of Hopf-links.

Below we specialize to a 4D nodal circuit with minimally nontrivial boundary linkage,
{termed ``2-link" below to distinguish from a Hopf link [Fig.~\ref{lattice_circuit1}].
{Following Eq. (\ref{h4DN})}, the circuit Laplacian is given by
\begin{eqnarray}
J_{\rm 4D}({\bm k})&=&[h_{1,\rm 3D}({\bm k}_{\rm 3D})(\cos k_w-1)\nonumber\\
&&+h_{2,\rm 3D}({\bm k}_{\rm 3D})(\cos k_w+1)+\sin k_w\sigma_3]\,i\tau_2,
\label{circuit_link2}
\end{eqnarray}
with $\sigma_i$ and $\tau_i$ the Pauli matrices acting on two different pseudospin degrees of freedom, and
\begin{eqnarray}
h_{1,\rm 3D}({\bm k}_{\rm 3D})&=&(\sin{k_y})\sigma_1\nonumber\\
&+&(m-\cos{k_x}-\cos{k_y}-\cos{k_z})\sigma_2;\label{pre-eq91} \\
h_{2,\rm 3D}({\bm k}_{\rm 3D})&=&-(\sin{k_z})\sigma_1\nonumber\\
&-&(m+\cos{k_x}-\cos{k_y}-\cos{k_z})\sigma_2.\label{pre-eq92}
\end{eqnarray}
Here $h_{1,\rm 3D}$ and $h_{2,\rm 3D}$ give mutually displaced NLs along the $k_x$-$k_z$ and $k_x$-$k_y$ planes respectively,
and they are manifested in the 3D surface Brillouin zone of the 4D system, in the same manner of Fig. \ref{fig1}(b) and (e).
Note that the tensor product with $\tau_2$ acts on every term, thus does not affect the NL structure of the system. However, it is necessary to have it in our circuit construction for maintaining time-reversal symmetry.}
In terms of real-space lattice circuit connections, this construction corresponds to capacitive/inductive elements for positive/negative couplings respectively (details in Fig.~\ref{lattice_circuit1}(a,b) and Supplementary Materials. In particular, Fig.~\ref{lattice_circuit1}(a) depicts a supercell with two unit cells in it and all the intral-supercell hoppings, and
Fig.~\ref{lattice_circuit1}(b) displays only the hoppings between different supercells.
For $1<m<2$, the two loops are linked, as shown in Fig.~\ref{lattice_circuit1}(c).

We now re-examine Kirchhoff's law in a general circuit context, and explain how the Seifert surfaces (extensive zero eigenvalues of the Laplacian) show up in impedance measurements. For a 4D circuit with OBCs in the $\hat w$ direction and PBCs in the other ${\bm k}_{\rm 3D}$ directions, Kirchhoff's law is expressed explicitly in terms of the boundary momentum ${\bm k}_{\rm 3D}$ as
\begin{equation}
I_a({\bm k}_{\rm 3D})=\sum_{b,{\bm k}'_{\rm 3D}}J_{ab}({\bm k}_{\rm 3D},{\bm k}'_{\rm 3D})V_b({\bm k}'_{\rm 3D}),
\label{IJV}
\end{equation}
where components of $I_a({\bm k}_{\rm 3D})$, $V_a({\bm k}_{\rm 3D})$ represent the ${\bm k}_{\rm 3D}$-th intra-layer Fourier component of the input current and electrical potential in layer $a$. Here the ``layers'' are 3D sublattices parallel to the open boundary, which collectively make up the 4D circuit. Anticipating disorder, we have not assumed that $J_{ab}$ is diagonal in momentum (translation invariant).

The key reason why our Seifert surfaces are so easily detectable is that they represent extensive degeneracies which manifest as ``topolectrical'' resonances~\cite{circuit3}. Consider a multi-terminal impedance measurement on a configuration with input currents $I_{b,r_1},I_{b,r_2},...$ 
into nodes {$(r_1, r_2, \cdots)$} at 3D layer $b$. We measure the potentials $V_{a,r_1},V_{a,r_2},...$ 
at nodes of layer $a$, which is not necessarily the same as $b$. From Eq.~\ref{IJV}, the potential and current Fourier components are related via
\medmuskip=0mu
\thickmuskip=0mu
\begin{eqnarray}
V_a({\bm k}_{\rm 3D})&=&
\sum_{b,{\bm k}'_{\rm 3D}}(J^{-1})_{ab}({\bm k}_{\rm 3D},{\bm k}'_{\rm 3D})I_b({\bm k}'_{\rm 3D})\notag\\
&\approx & \sum_{n;b}\frac{\left[|\psi_n({\bm k}_{\rm 3D})\rangle\langle\psi_n({\bm k}_{\rm 3D})|\right]_{ab}}{j_n({\bm k}_{\rm 3D})}I_b({\bm k}_{\rm 3D})\notag\\
&=&\sum_{b}Z_{ab}({\bm k}_{\rm 3D})I_b({\bm k}_{\rm 3D})
\label{IJV2}
\end{eqnarray}
with $Z_{ab}({\bm k}_{\rm 3D})$ the ${\bm k}_{\rm 3D}$ wavevector impedance, and $j_n$ and $|\psi_n\rangle$ the $n$-th eigenvalue and eigenvector of the circuit Laplacian $J$, expressed in the $(a,{\bm k}_{\rm 3D})$ basis above. The crucial observation is that $V_a({\bm k}_{\rm 3D})$ is expected to \textit{diverge} when an extensive number of zero modes (with $j_n\approx 0$) are present. In our context, the divergence of $V_a({\bm k}_{\rm 3D})$ indicates a Seifert surface state at ${\bm k}_{\rm 3D}$ when $a$ is the surface 3D layer; a similar though weaker divergence in the bulk will indicate a bulk nodal crossing at ${\bm k}_{\rm 3D}$.

To probe the Seifert surfaces, we simulate an experiment where currents enter nodes in unit cell layer $b=1,2,$ or $3$, layer $1$ being the 3D boundary, with input current magnitudes modulated according to the ${\bm k}_{\rm 3D}$ momentum wavevector. This is consistent with overall current conservation as long as ${\bm k}_{\rm 3D}\,\neq\, \bold 0$. Next, we take the simulated voltage readings on nodes in layers $a=1,2,$ and $3$, and extract their ${\bm k}_{\rm 3D}$-th Fourier component. The results for $\log |Z_{ab}({\bm k}_{\rm 3D})|$ for the model of Eq. (\ref{circuit_link2}) is shown in Fig.~\ref{lattice_circuit1}(d-f) for $1\%$ disorder [see the Supplementary Materials \cite{SuppMat}].

As evident in Figs.~\ref{lattice_circuit1}(c-d), we clearly observe a Seifert surface as pronounced resonance peaks at the boundary $(a,b)=(1,1)$. These resonances gradually decay as {the layer index under measurement moves towards the bulk} [Figs.~\ref{lattice_circuit1}(e-f)], eventually morphing into the bulk NLs.

\subsection{Topological classification through Seifert surfaces} 
\begin{figure}[h]
\includegraphics[width=1\linewidth]{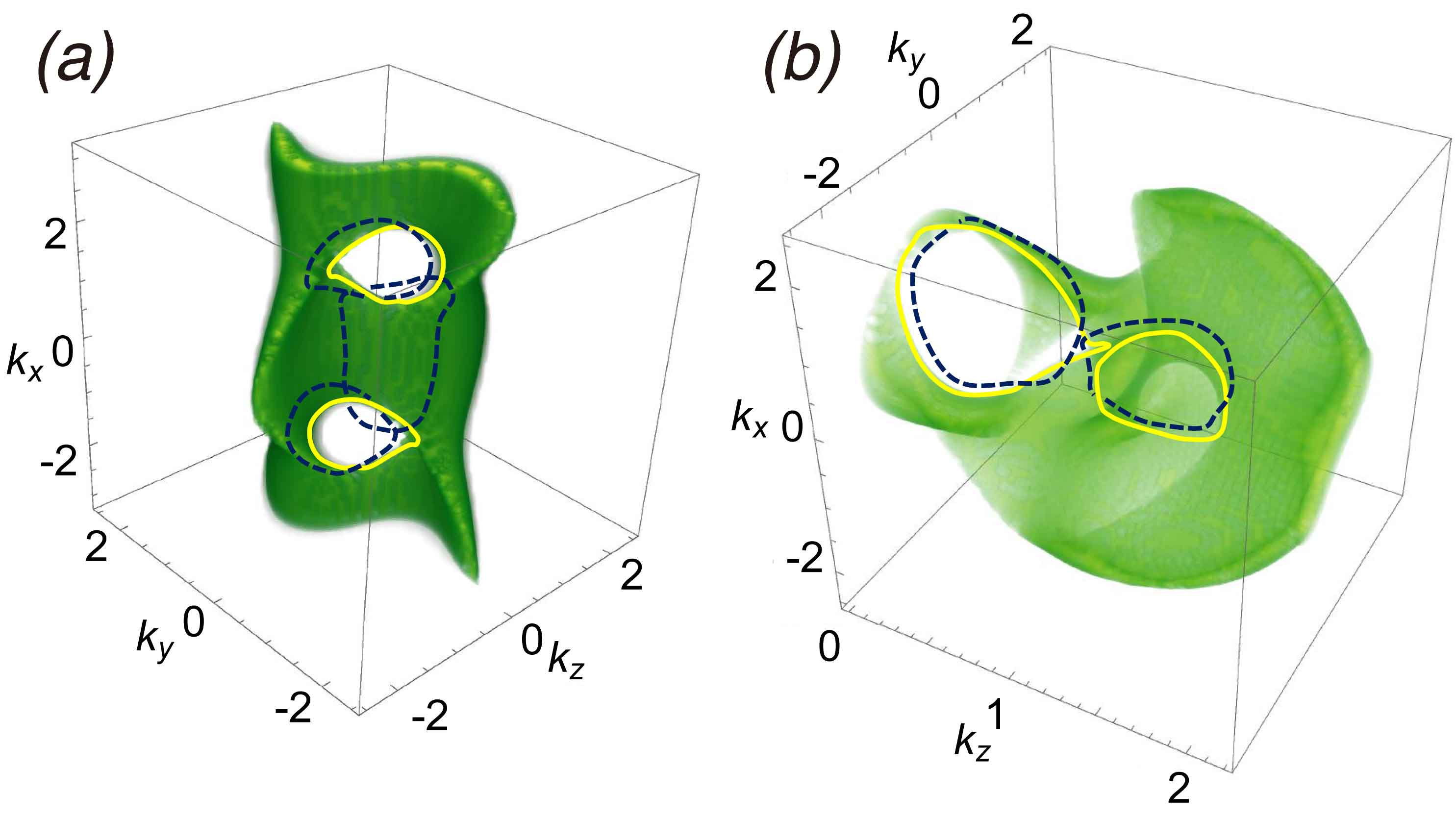}
\caption{Basis homology (yellow) and lifted homology (dashed blue) loops of the Seifert surfaces of the 2-link (a) and Trefoil (b) NL systems reconstructed from topolectrical resonances illustrated in Fig.~\ref{lattice_circuit1}). There are respectively $R=3$ and $R=2$ homology bases of each type. Linking numbers between the yellow and dashed blue loops yield the Seifert matrix elements displayed in Table.~\ref{table1}. For clarity, loops with vanishing linkages are omitted.}
\label{lattice_circuit4}
\end{figure}

Although the Seifert surface obtained is not the unique surface bounded by the NLs, valuable nodal topology information can nevertheless be extracted. Most obvious is the number of components (loops) $N$ in the nodal structure, which corresponds to the number of punctures in the Seifert surface. Mathematically capping them with disks, the resultant Seifert surface becomes a closed Riemann surface with genus $g$ handles. Although this genus is somewhat hard to directly visualize due to the intricate shape of the Seifert surface [see for instance Fig.~\ref{lattice_circuit4}, both with genus $1$], it can be systematically computed by probing the connectivity of the the zero mode manifold as described below. The minimal $g$ for a given NL structure is also a topological invariant.

More sophisticated invariants are encoded in the homology properties of the Seifert surface, as captured by the Seifert matrix $S$ of linking numbers between its homology generators and those of its lifted (infinitesimally shifted) counterpart. The latter can be obtained by perturbing the coefficient of {the system}
by a small real constant, which is easily implementable in circuits via a small AC frequency shift. Due to the robustness of the topology of the Seifert surface, we emphasize that the same Seifert matrix will be obtained {regardless} of the choice of the small frequency shift, as long as the same shift is consistently used in the measurements. Shown in Fig.~\ref{lattice_circuit4} are illustrative homology (yellow) and lifted homology (dashed blue) generators from topolectrical resonant Seifert surfaces of 2-link (Fig.~\ref{lattice_circuit1}) and Trefoil knots [see Supplementary Materials], whose linking numbers are summarized in the Seifert matrices of Table~\ref{table1}. For NLs with simply connected Seifert islands, the number of homology basis generators (rank) is given by $R =2g+N-1$. From them, NL knot invariants like the Alexander polynomial $A(t)=t^{-R/2}\text{Det}\,(S-tS^T)$ and knot Signature (\# positive - \# negative eigenvalues of $S$) can be extracted (Table.~\ref{table1}). 

\begin{table}[h]
\begin{minipage}{\linewidth}
\centering
\renewcommand{\arraystretch}{2}
\resizebox{\linewidth}{60pt}
{
\begin{tabular}{|c|c|c|c|c|c|c|}\hline
\bf{Link/Knot} &\bf{N}& \bf{Genus}&\bf{Rank}&\bf{Seifert matrix} & \bf{$A(t)$}  & \bf{Signature}\\    \hline
\bf{Hopf} & $2$ & $0$ & $1$ & $-1$ & $\frac{t-1}{\sqrt{t}}$  & $-1$\\ \hline
\bf{Trefoil} & $1$ & $1$ & $2$ & $\left(\begin{matrix}-1 & 0 \\ 1 & -1\end{matrix}\right)$ & $t+t^{-1}-1$  & $-2$\\ \hline
\bf{2-link} & $2$ & $1$ & $3$ & $\left(\begin{matrix}-1 & 0&0 \\ 1 & 0 & -1 \\ 0 & 0 & 1\end{matrix}\right)$ & $0$  & $0$\\ \hline
\end{tabular}
}
\end{minipage}
\bigskip
\caption{Various Seifert surface properties and NL topological invariants from simulated boundary Seifert surface measurements. }
\label{table1}
\end{table}

\section{Discussion}
{\it Seifert surface} is now elevated in this work from a sophisticated mathematical concept to an experimentally accessible object
crucial for topological characterization. This work also discovers a simple means to realize rather arbitrary
 linkage and knot topology of momentum-space NLs using an effective 4D space, but still using practical experimental settings in 3D.
 That is,  the introduction of a 4th dimension in our approach presents no additional practical difficulties in circuit realizations.

The Seifert surfaces that can now be fully imaged encode \textit{full} 3D nodal structure information, superior to usual 2D drumhead surface states whereby detailed 3D geometric and topological information is irretrievably lost through surface ``shadows" projections.
With their existence rooted in 2D Chern topology, NLs/knots and their Seifert surface states are 4D analogs of 3D Weyl points and Fermi arcs. Yet, intricacies of their higher dimensional structure far transcend any characterization by a single Chern number.
In the several explicit models discussed here and in the Supplementary Materials \cite{SuppMat}, the lattice couplings are carefully designed to give clear and quantitative illustrations of the nontrivial NL topology through the topological invariants extracted from their topolectrically resonant Seifert surfaces.
In this regard, RLC circuit setups are the most suitable experimental platform, as couplings can be simulated by independently tuned capacitors and inductors.
Besides circuits, this work provides a potential scheme to realize exotic links and knots together with their Seifert surfaces in quantum systems, such as optical cold-atom lattices with one or more synthetic dimensions~\cite{synthetic_dimension1,synthetic_dimension2,synthetic_dimension3,synthetic_dimension4}.



\twocolumngrid
\section{Acknowledgments}
L.L. and C.H.L contributed equally to this work.  J.G. thanks Prof. Xiangang Wan for thought-provoking discussions on nodal-line semimetals.   J.G. acknowledges research funding by the Singapore NRF grant No. NRF-NRFI2017-04 (WBS No. R-144-000-378-281).

\clearpage
\onecolumngrid
\begin{center}
\textbf{\large Supplementary Materials}\end{center}
\setcounter{equation}{0}
\setcounter{figure}{0}
\renewcommand{\theequation}{S\arabic{equation}}
\renewcommand{\thefigure}{S\arabic{figure}}

\section{Protection of 4D NLs by 2D Chern invariants}\label{app:Chern}
{A 4D 2-band system does not require any symmetry protection to host 1D nodal lines. To comprehend this, we view the 4D system as a 2D quantum Hall system described by a Chern number, with other two momenta taken as additional system parameters. For instance, the bulk Hamiltonian in the 4D momentum space of $(k_x,k_y,k_z,k_w)$
can be viewed as a 2D Hamiltonian in $(\hat{z},\hat{w})$ dimensions, with $(k_x,k_y)$ being two system parameters.
If the band Chern number of $h_{\rm 4D}({\bm k})$ with respect to $(k_z,k_w)$ is nonzero, then with OBCs along the $\hat{w}$ dimension, chiral boundary states must emerge, as protected by the Chern invariant.
In the absence of a diagonal term $h_0\mathbb{I}$ in the spinor representation, the chiral boundary states become degenerate at zero-energy at some symmetric points of $k_z$, which then yields the Seifert Fermi (zero energy) surface when we scan $k_x$ and $k_y$.

As a simple example, consider a minimal 4D system with a single NL, described by the following Hamiltonian,
\begin{eqnarray}
h_{\rm 4D}({\bm k})=(\cos{k_x}+\cos{k_y}+\cos{k_z}+\cos{k_w}-m)\sigma_a+(\sin{k_w})\sigma_c+(\sin{k_z})\sigma_b, \label{single_NL}
\end{eqnarray}
with $\sigma_{a,b,c}$ an arbitrary permutation of the three Pauli matrices.
When $2<m<4$, this system has a single NL in the $k_x-k_y$ plane with $k_w=k_z=0$, as shown in Fig.~\ref{figS1}(a).
Fig.~\ref{figS1}(b)-(d) display the spectrum versus $k_z$ with $k_{y}=0$ and several representative values of $k_x$, under OBCs along the $\hat{w}$ direction.  The zero-energy chiral states are located at $k_z=0$. Taking $k_x$ and $k_y$ as two parameters, at $k_y=0$, $\cos k_x=0$ represents the topological phase transition point of the 2D quantum Hall system, as shown in Fig.~\ref{figS1}(b)-(d). The chiral edge states exist only for $\cos(k_x)>0$ but not for  $\cos(k_x)<0$.
With $k_x$ and $k_y$ continuously varying, these zero-energy boundary states form a simple 2D Seifert surface matching the NL shown in  Fig.~\ref{figS1}(a).

\begin{figure}
\includegraphics[width=1\linewidth]{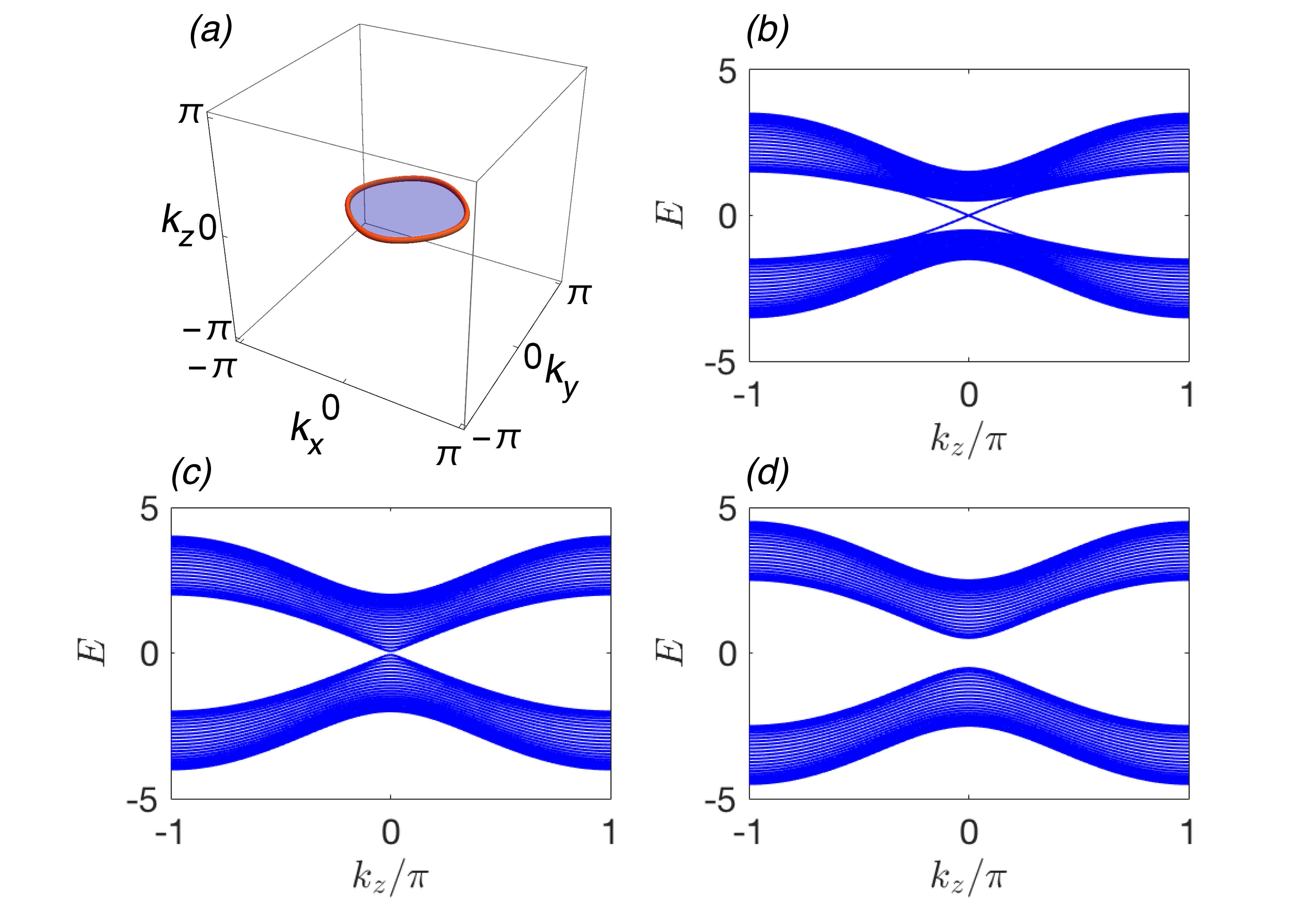}
\caption{
A single NL of the Hamiltonian in Eq.~\ref{single_NL} and its spectra under OBCs along the $\hat{w}$ direction, with $k_x$ and $k_y$ viewed as two system parameters.  (a) The NL (red loop) and the zero-energy boundary states (blue plane). (b)-(d) Spectrum versus $k_z$, with $k_y=0$, (b) $k_x=\pi/3$, (c) $k_x=\pi/2$, and (d) $k_x=2\pi/3$. Gapless chiral boundary states seen in panel (b) are protected by a nonzero Chern invariant as in standard 2D quantum Hall effect. In this simple model, zero-energy boundary states are obtained at $k_z=0$.}
\label{figS1}
\end{figure}

To further understand when this 2D system is topologically nontrivial, we can rewrite the Hamiltonian as
\begin{eqnarray}
h_{\rm 4D}({\bm k})= \left[h_a({\bm k}_{\rm 3D})+(\cos{k_w}-1)\right]\sigma_a+(\sin{k_w})\sigma_c+h_b({\bm k}_{\rm 3D})\sigma_b,
\end{eqnarray}
with
$h_a({\bm k}_{\rm 3D})= \cos{k_x}+\cos{k_y}+\cos{k_z}-2$ and $h_b({\bm k}_{\rm 3D})=\sin{k_z}$, and view
the system as a 1D superlattice along the $\hat{w}$ direction, with $(k_x, k_y, k_z)$ serving as three system parameters. With this perspective, $h_b({\bm k}_{\rm 3D})$ can be understood as a mass term, which must be zero to yield zero-energy boundary states {protected by a chiral symmetry $\sigma_bh_{\rm 4D}({\bm k})\sigma_b=-h_{\rm 4D}({\bm k})$}.
The winding of the vector $[h_a({\bm k}_{\rm 3D})+(\cos{k_w}-1), \sin{k_w}]$ around the origin
as $k_w$ varies from $0$ to $2\pi$ determines if there are boundary states
upon taking OBCs along the $\hat{w}$ direction.  A nonzero winding of this kind requires $2>h_a({\bm k}_{\rm 3D})>0$, which guarantee topological zero-energy boundary states under the symmetric condition $h_b({\bm k}_{\rm 3D})=0$. Thus topological phase transitions occur at $h_b({\bm k}_{\rm 3D})=0$ and $h_a({\bm k}_{\rm 3D})=0$ or $2$, which gives the NL as the bulk gap closes here at zero energy.
Therefore, the collection of the zero-energy boundary states must be the Seifert surface matching the NL structure because these boundary states are obtained with $2>h_a({\bm k}_{\rm 3D})>0$ and $h_b({\bm k}_{\rm 3D})=0$.

\section{An explicit 4D model with a nodal Trefoil knot}
With $h_{\rm 4D}({\bf k})$ described by Eq. 4 and Eq. 5 of the main text, the following example
\begin{eqnarray}
h_a({\bf k}_{\rm 3D})&=&[\sin^3 k_z-3\sin k_z\sin^2 k_x+\sin^2 k_y-(\cos k_z+\cos k_y+\cos k_x-m)^2],\\
h_b({\bf k}_{\rm 3D})&=&[-3\sin^2 k_z \sin k_x + \sin^3 k_x +  2 \sin k_y (\cos k_z+\cos k_y+\cos k_x-m)],
\end{eqnarray}
with $t_w=2$ and $m=2$ yields the Trefoil knot model in Fig.~2(d) of the main text. The above $h_{a,b}({\bf k}_{\rm 3D})$ are obtained from
\begin{eqnarray}
z^3+w^2=h_a({\bf k}_{\rm 3D})+ ih_b({\bf k}_{\rm 3D}),
\end{eqnarray}
with the regularized stereographic map
\begin{eqnarray}
z&=&\sin k_z-i\sin k_x,\nonumber\\
w&=&\sin k_y+i(\cos k_z+\cos k_y+\cos k_x-m).
\end{eqnarray}
 More general constructions of $h({\bf k}_{\rm 3D})$ to obtain other nodal knots or links can obtained from various methods, e.g. the Hopf map indexed with a pair of numbers $(p,q)$~\cite{linkknot2_SI}.  For the Trefoil knot constructed above, $(p,q)=(3,2)$.

\section{Boundary states featuring arbitrarily many linked NLs}
Here we provide an explicit ansatz to construct 4D models whose boundary states can accommodate arbitrarily many linked NLs.
Consider the Hamiltonian
\begin{eqnarray}
h_{N,{\rm 4D}}({\bf k})=\sum_{n=1}^{2M}h_{n,{\rm 3D}}({\bf k}_{\rm 3D}) g_n(k_w)+f(k_w)\sigma_c,
\end{eqnarray}
where each individual $h_{n,{\rm 3D}}({\bf k}_{\rm 3D})$ contains only two Pauli matrices $\sigma_{a,b}$ and may describe a 3D NL system. The condition for this 4D Hamiltonian to yield $2M$ NLs, each lying in a 3D slice with a different $k_w$ properly, is already specified in the main text.
For the sake of presenting an explicit models here and a better demonstration of the numerical results of Seifert surfaces in Fig.~2(e,f) of the main text, we adopt a slightly different construction here.
We first define a function $f(k_w)$ as
\begin{eqnarray}
f(k_w)=\prod_{n=1}^{M} \sin (k_w-\alpha_n),
\end{eqnarray}
where $\alpha_n$ is chosen as $0<\alpha_n<\pi$, $\alpha_n\neq\pi/2$, and $\alpha_i\neq\alpha_j$ for $i\neq j$. This function has zeros at $k_{w,n}=\alpha_n$ and $k_{w,(M+n)}=\alpha_n+\pi$, with in total $2M$ different solutions to $f(k_w)=0$.
We next consider the following explicit $g_n(k_w)$,
\begin{eqnarray}
g_n(k_w)=(\cos k_w+\cos{{k}}_{w,n})\prod_{n'\neq n,~n'\neq n-M}^{M}\sin(k_w-\alpha_{n'}),\nonumber\\
\end{eqnarray}
which equals to zero at any $k_w=k_{w,m}$ except for $m=n$.
Thus the total Hamiltonian
\begin{eqnarray}
h_{N,{\rm 4D}}({\bf k})=\sum_{n=1}^{2M}h_{n,{\rm 3D}}({\bf k}_{\rm 3D}) g_n(k_w)+f(k_w)\sigma_c
\end{eqnarray}
can host up to $2M$ nodal loops. To have $N \leqslant 2M$ NLs in this system,
we require each $h_{n,{\rm 3D}}({\bf k}_{\rm 3D})$ describes a 3D single-NL system for $n\in[1,N]$, and a 3D insulating system for $N<n\leqslant 2M$. For simplicity, here we choose $h_{n,{\rm 3D}}({\bf k}_{\rm 3D})=\sigma_a+\sigma_b$ for all $N<n\leqslant 2M$. Therefore the system $h_{N,{\rm 4D}}({\bf k})$ constructed above has $N$ NLs, each given by a $h_{n,{\rm 3D}}({\bf k}_{\rm 3D})$ at $k_w=k_{w,n}$ with $n\leqslant N$.


The above construction leads to both specific examples discussed in the main text. The first example is the Hopf-link, which is obtained by choosing $M=1$ and $\alpha_1=0$. The Hamiltonian is given by
\begin{eqnarray}
h_{\rm 4D}({\bf k})&=&h_{1,{\rm 3D}}({\bf k}_{\rm 3D})(\cos k_w+1)+h_{2,{\rm 3D}}({\bf k}_{\rm 3D})(\cos k_w-1)+(\sin k_w) \sigma_c,
\end{eqnarray}
with 
\begin{eqnarray}
h_{1,{\rm 3D}}({\bf k}_{\rm 3D})&=&[m-\cos{(k_x+\alpha)}-\cos{k_y}-\cos{k_z}]\sigma_a+(\sin{k_y})\sigma_b;\nonumber\\
h_{2,{\rm 3D}}({\bf k}_{\rm 3D})&=&-[m-\cos{(k_x-\alpha)}-\cos{k_y}-\cos{k_z}]\sigma_a-(\sin{k_z})\sigma_b.\nonumber
\end{eqnarray}
 For $1<m<3$ and $\alpha=0$, $h_{1,{\rm 3D}}({\bf k}_{\rm 3D})$ and $h_{2,{\rm 3D}}({\bf k}_{\rm 3D})$ give two nodal loops both centering at $(k_x,k_y,k_z)=(0,0,0)$. A nonzero $\alpha$ shifts the two NLs along $k_x$ in opposite directions.
When OBC is taken along the $\hat{w}$ direction, we obtain a pair of Hopf-link NLs in the 3D parameter space of ${\bf k}_{{\rm 3D}}$, and the boundary Fermi (zero-energy) surface gives the Seifert surface of the Hopf-link, as shown in Fig.~2(e) of the main text with $m=2$ and
{$\alpha=\pi/4$}.

The second example is a set of Borromean rings, which are three NLs linked together but any two of them are not linked. This is obtained by choosing $M=2$, $\alpha_1=0$, $\alpha_2=\pi/4$, with
\begin{eqnarray}
h_{1,{\rm 3D}}({\bf k}_{\rm 3D})&=&(m-\cos k_x-B\cos k_y-A\cos k_z)\sigma_a+\sin k_x\sigma_b;\nonumber\\
h_{2,{\rm 3D}}({\bf k}_{\rm 3D})&=&-(m-A\cos k_x-\cos k_y-B\cos k_z)\sigma_a-\sin k_y\sigma_b;\nonumber\\
h_{3,{\rm 3D}}({\bf k}_{\rm 3D})&=&(m-B\cos k_x-A\cos k_y-\cos k_z)\sigma_a+\sin k_z\sigma_b,\nonumber
\end{eqnarray}
and $h_{4,{\rm 3D}}^a=h_{4,{\rm 3D}}^b=1$. The 4D Hamiltonian is then given by
\begin{eqnarray}
h_{\rm 4D}({\bf k})&=&h_{1,{\rm 3D}}({\bf k}_{\rm 3D})(\cos k_w+1)\sin(k_w-\pi/4)+h_{2,{\rm 3D}}({\bf k}_{\rm 3D})(\cos k_w+\sqrt{2}/2)\sin k_w\nonumber\\
&&+h_{3,{\rm 3D}}({\bf k}_{\rm 3D})(\cos k_w-1)\sin(k_w-\pi/4)+h_{4,{\rm 3D}}({\bf k}_{\rm 3D})(\cos k_w-\sqrt{2}/2)\sin k_w\nonumber\\
&&+\sin k_w\sin (k_w-\pi/4)\sigma_c.
\end{eqnarray}
The coefficients $A$ and $B$ are to stretch the loops in different directions. Fig.~2(f) of the main text has shown the Borromean rings and the boundary Fermi states with $m=2$, $A=1.2$ and $B=0.6$.

\section{Circuit realizations of nodal ring and Hopf-link 4D NLs}\label{app:circuit_loop}
A minimal model of NL circuit system in 4D can be described by the circuit Laplacian
\begin{eqnarray}
J_{\rm min}({\bm k})=i[(m-\cos{k_x}-\cos{k_y}-\cos{k_w})\sigma_1+\sin{k_w}\sigma_2+\cos{k_z}\sigma_3],\label{minimal_circuit}
\end{eqnarray}
which gives two parallel NLs with $(k_z,k_w)=(\pm\pi/2,0)$ when $1<m<3$. The Seifert surfaces of these NLs are simply two isolated disks and are topologically trivial, as shown in Fig.~\ref{loops_circuit}(a).

\begin{figure}
\includegraphics[width=0.8\linewidth]{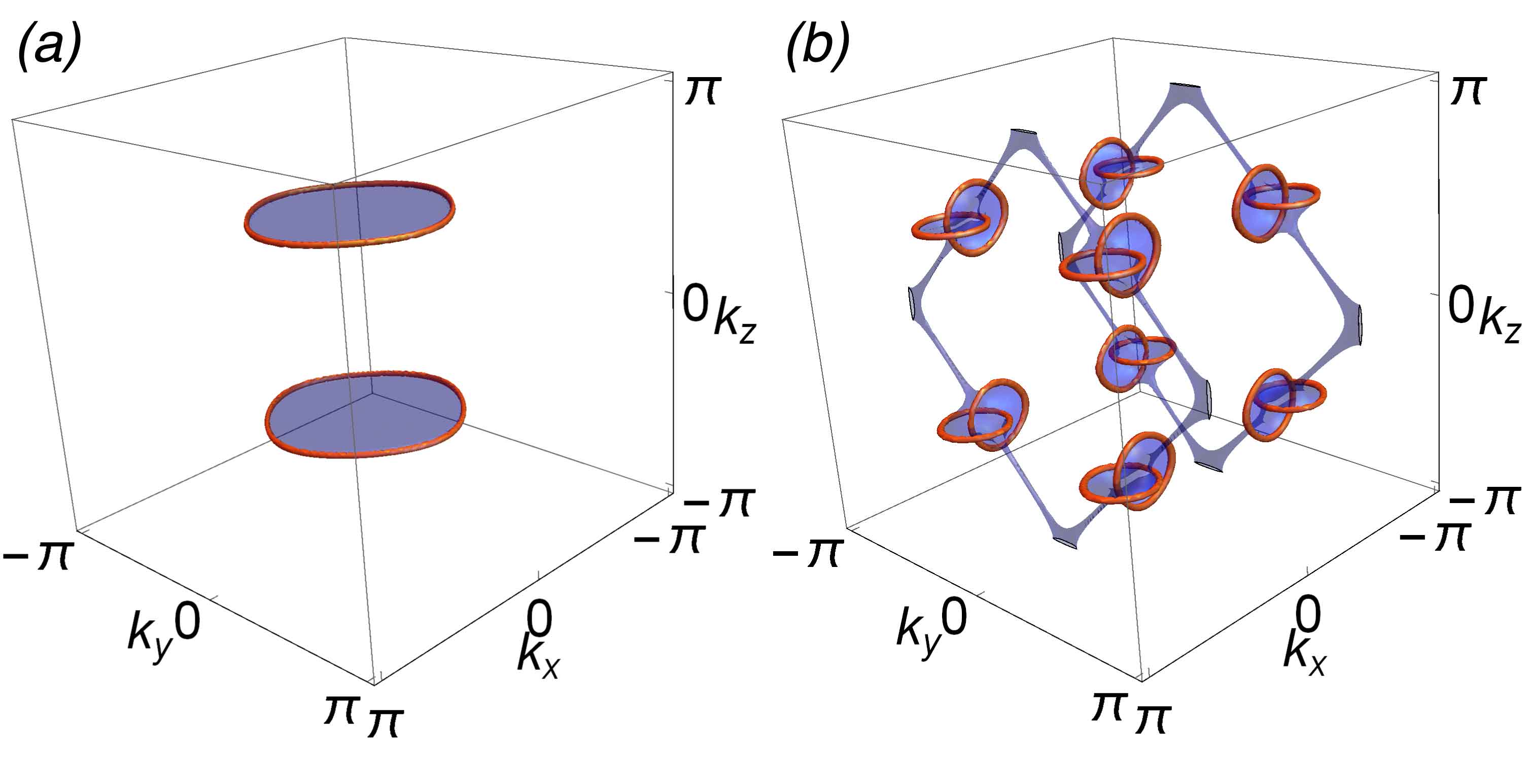}
\caption{The NLs (red loops) and the boundary Fermi surface (blue regions) of the systems given by (a) Eq.~\ref{minimal_circuit} with $m=2$, and (b) Eq.~\ref{minimal_circuit2} with $m=0.5$ and $\mu=0.2$, respectively. The ``tube-like" regions of the boundary Fermi Seifert surface (right panel) connect the surface regions around different copies of the Hopf-links.}
\label{loops_circuit}
\end{figure}

While a Hamiltonian realizing a true Hopf-link cannot be time-reversal symmetric by itself, it is still possible to design circuits that yield pairs of Hopf-links, which are time-reversal symmetric to each other. Consider a second circuit Laplacian (which is not used in the main text or in our following simulations),
\begin{eqnarray}
J({\bm k})=i[h_{1,{\rm 3D}}({\bm k}_{\rm 3D}) (\cos k_w-1)+ h_{2,{\rm 3D}}({\bm k}_{\rm 3D}) (\cos k_w+1)+\sin k_w\sigma_2\tau_1], \label{minimal_circuit2}
\end{eqnarray}
where  $h_{1,{\rm 3D}}({\bm k}_{\rm 3D})$ and $h_{2,{\rm 3D}}({\bm k}_{\rm 3D})$ are the 3D Hamiltonians with NLs in different planes, given by
\begin{eqnarray}
h_{1,{\rm 3D}}({\bm k}_{\rm 3D}) &=&(\cos k_x+\mu)\sigma_1+\cos k_y\sigma_2\tau_2+\cos k_z\sigma_3+m\sigma_1\tau_1;\nonumber\\
h_{2,{\rm 3D}}({\bm k}_{\rm 3D})&=&-(\cos k_x-\mu)\sigma_1-\cos k_z\sigma_2\tau_2-\cos k_y\sigma_3+m\sigma_1\tau_1.\nonumber
\label{circuit_link3}
\end{eqnarray}
Each of $h_1({\bm k}_{\rm 3D})$ and $h_2({\bm k}_{\rm 3D})$ describes a NL system with eight NLs lying in $x-y$ and $x-z$ planes respectively, and each NL carries a nonzero monopole charge~\cite{Z2_loops_SI}. The radii of these loops are determined by $m$, and the centers of these loops are at ${\bm k}_{\rm 3D}=(\arccos (-\mu),\pm\pi/2,\pm\pi/2)$ for $h_{1,{\rm 3D}}({\bm k}_{\rm 3D})$, and at ${\bm k}_{\rm 3D}=(\arccos (\mu),\pm\pi/2,\pm\pi/2)$ for $h_{2,\rm{3D}}({\bm k}_{\rm 3D})$. The loops of these two 3D Hamiltonians are linked together when $\mu$ is small but nonzero. Furthermore, the model satisfies reflection symmetries $J(k_i)=J(-k_i)$ for $i=x,y,z$, thus the eight Hopf-links are symmetric to each other.
The energy dispersion of this model is given by
\begin{eqnarray}
E=\pm\sqrt{(\sqrt{P_2^2+P_2'^2}\pm m)^2+Q_2^2+\sin^2 k_w},
\end{eqnarray}
with $P_2=2\mu\cos k_w-2\cos k_x$, $P_2'=\cos k_y(\cos k_w-1)-\cos k_z(\cos k_w+1)$ and $Q_2=\cos k_z(\cos k_w-1)-\cos k_y(\cos k_w+1)$.
In Fig.~\ref{loops_circuit}(b) we show 
the eight Hopf-links with $m=0.5$ and $\mu=0.2$. The boundary flatband connects the pair of loops, and also extends to the other pairs.

\section{Details of the 2-link and Trefoil circuits}\label{app:circuit_knot}

\subsection{Details of the 2-link circuit}
The lattice structure of the 2-link circuit in our simulations can be obtained from an inverse Fourier transformation of the circuit Laplacian ($J_{\rm 4D}({\bm k})$ for Eq.~9 of the main text), which takes the following lattice form
\begin{eqnarray}
-iJ_{\rm 4D}&=&\sum_n 2m(\hat{a}^{\dagger}_{\uparrow,n}\hat{b}_{\downarrow,n}-\hat{a}^{\dagger}_{\downarrow,n}\hat{b}_{\uparrow,n})+h.c.\nonumber\\
&-&\sum_n
\frac{1}{2}\left[\hat{a}^{\dagger}_{\uparrow,n}\hat{b}_{\uparrow,n+\hat{w}}-\hat{a}^{\dagger}_{\uparrow,n}\hat{b}_{\uparrow,n-\hat{w}}
-\hat{a}^{\dagger}_{\downarrow,n}\hat{b}_{\downarrow,n+\hat{w}}+\hat{a}^{\dagger}_{\downarrow,n}\hat{b}_{\downarrow,n-\hat{w}}
\right]+h.c.\nonumber\\
&+&\sum_n \left[-\frac{1}{2}\hat{a}^{\dagger}_{\uparrow,n}\hat{b}_{\downarrow,n+\hat{y}}-\frac{3}{2}\hat{a}^{\dagger}_{\uparrow,n}\hat{b}_{\downarrow,n-\hat{y}}
+\frac{3}{2}\hat{a}^{\dagger}_{\downarrow,n}\hat{b}_{\uparrow,n+\hat{y}}+\frac{1}{2}\hat{a}^{\dagger}_{\downarrow,n}\hat{b}_{\uparrow,n-\hat{y}}\right]+h.c.\nonumber\\
&+&\sum_n \left[-\frac{1}{2}\hat{a}^{\dagger}_{\uparrow,n}\hat{b}_{\downarrow,n+\hat{z}}-\frac{3}{2}\hat{a}^{\dagger}_{\uparrow,n}\hat{b}_{\downarrow,n-\hat{z}}
+\frac{3}{2}\hat{a}^{\dagger}_{\downarrow,n}\hat{b}_{\uparrow,n+\hat{z}}+\frac{1}{2}\hat{a}^{\dagger}_{\downarrow,n}\hat{b}_{\uparrow,n-\hat{z}}\right]+h.c.\nonumber\\
&+&\sum_n \frac{1}{2}\left[\hat{a}^{\dagger}_{\uparrow,n}\hat{b}_{\downarrow,n+\hat{x}+\hat{w}}+\hat{a}^{\dagger}_{\uparrow,n}\hat{b}_{\downarrow,n+\hat{x}-\hat{w}}
+\hat{a}^{\dagger}_{\uparrow,n}\hat{b}_{\downarrow,n-\hat{x}+\hat{w}}+\hat{a}^{\dagger}_{\uparrow,n}\hat{b}_{\downarrow,n-\hat{x}-\hat{w}}\right]+h.c.\nonumber\\
&-&\sum_n \frac{1}{2}\left[\hat{a}^{\dagger}_{\downarrow,n}\hat{b}_{\uparrow,n+\hat{x}+\hat{w}}+\hat{a}^{\dagger}_{\downarrow,n}\hat{b}_{\uparrow,n+\hat{x}-\hat{w}}
+\hat{a}^{\dagger}_{\downarrow,n}\hat{b}_{\uparrow,n-\hat{x}+\hat{w}}+\hat{a}^{\dagger}_{\downarrow,n}\hat{b}_{\uparrow,n-\hat{x}-\hat{w}}\right]+h.c.\nonumber\\
&-&\sum_n
\frac{1}{4}\left[\hat{a}^{\dagger}_{\uparrow,n}\hat{b}_{\downarrow,n+\hat{y}+\hat{w}}+\hat{a}^{\dagger}_{\uparrow,n}\hat{b}_{\downarrow,n+\hat{y}-\hat{w}}
-\hat{a}^{\dagger}_{\uparrow,n}\hat{b}_{\downarrow,n-\hat{y}+\hat{w}}-\hat{a}^{\dagger}_{\uparrow,n}\hat{b}_{\downarrow,n-\hat{y}-\hat{w}}\right]+h.c.\nonumber\\
&-&\sum_n
\frac{1}{4}\left[\hat{a}^{\dagger}_{\downarrow,n}\hat{b}_{\uparrow,n+\hat{y}+\hat{w}}+\hat{a}^{\dagger}_{\downarrow,n}\hat{b}_{\uparrow,n+\hat{y}-\hat{w}}
-\hat{a}^{\dagger}_{\downarrow,n}\hat{b}_{\uparrow,n-\hat{y}+\hat{w}}-\hat{a}^{\dagger}_{\downarrow,n}\hat{b}_{\uparrow,n-\hat{y}-\hat{w}}\right]+h.c.\nonumber\\
&+&\sum_n
\frac{1}{4}\left[\hat{a}^{\dagger}_{\uparrow,n}\hat{b}_{\downarrow,n+\hat{z}+\hat{w}}+\hat{a}^{\dagger}_{\uparrow,n}\hat{b}_{\downarrow,n+\hat{z}-\hat{w}}
-\hat{a}^{\dagger}_{\uparrow,n}\hat{b}_{\downarrow,n-\hat{z}+\hat{w}}-\hat{a}^{\dagger}_{\uparrow,n}\hat{b}_{\downarrow,n-\hat{z}-\hat{w}}\right]+h.c.\nonumber\\
&+&\sum_n
\frac{1}{4}\left[\hat{a}^{\dagger}_{\downarrow,n}\hat{b}_{\uparrow,n+\hat{z}+\hat{w}}+\hat{a}^{\dagger}_{\downarrow,n}\hat{b}_{\uparrow,n+\hat{z}-\hat{w}}
-\hat{a}^{\dagger}_{\downarrow,n}\hat{b}_{\uparrow,n-\hat{z}+\hat{w}}-\hat{a}^{\dagger}_{\downarrow,n}\hat{b}_{\uparrow,n-\hat{z}-\hat{w}}\right]+h.c..
\label{circuit2link}
\end{eqnarray}
The basis is defined in Fig.~4 of the main text, with positive and negative couplings implemented by capacitors/inductors. The AC frequency $\omega$, capacitances $C$ and inductances $L$ should be chosen such that the ratio of their admittances $i\omega C/(i\omega L)^{-1}=-\omega^2 LC$ are of the appropriate proportions as indicated by Eq.~\ref{circuit2link}.

{When $1<m<3$, $h_{1,{\rm 3D}}({\bm k}_{\rm 3D})$ from Eq.~10 of the main text gives a NL in $k_x-k_z$ plane, centering at ${\bm k}_{\rm 3D}=(0,0,0)$; and $h_{2,{\rm 3D}}({\bm k}_{\rm 3D})$ from Eq.~11 of the main text} gives gives another NL in $k_x-k_y$ plane, centering at ${\bm k}_{\rm 3D}=(\pi,0,0)$. These two loops are linked to each other when $1<m<2$, as shown in Fig.~4(c) of the main text, with Seifert surface locally resembling that of a Hopf link.

The energy dispersion of this model is given by
\begin{eqnarray}
E=\pm\sqrt{P_1^2+Q_1^2+\sin^2 k_w},
\end{eqnarray}
with $P_1=(\sin k_y-\sin k_z)\cos k_w-(\sin k_y+\sin k_z)$ and $Q_1=2(\cos k_y +\cos k_z-m)-2\cos k_x\cos k_w$.
This Hamiltonian describes a 4-band model with two-fold degeneracy, and its energy dispersion is identical to the one of an analytically simpler 2-band model, described by Eq. 9 of the main text without the $\tau_2$. However, here we have to form a tensor product with $\tau_2$ in order to have back the time-reversal symmetry.

Due to the topologically non-trivial 3-torus BZ where this 2-link is embedded in, the latter becomes trivially linked, even though it is locally similar to the non-trivially linked Hopf-link. Nevertheless, the Seifert surface which interpolates the two NLs is highly non-trival, having a similar structure as that of a Hopf-link [Fig.~2(e) of the main text] locally near linkages, but stretching across the two NLs in a different way. This is reflected in the non-trivial form of its Seifert matrix, which contains more information than its Alexander polynomial which vanishes due to its trivial NL linkages (Table I of the main text).

\subsection{Details of the Trefoil knot circuit}
The Trefoil knot circuit used in the simulation presented in Fig.~5(b) of the main text is more complicated, and we shall just present its momentum-space structure. It can be numerically verified that $h_{\rm 3D}({\bm k}_{\rm 3D})=h_x({\bm k}_{\rm 3D})\sigma_x+h_z({\bm k}_{\rm 3D})\sigma_z$ gives, via Eqs.~3 to 5 of the main text, an RLC nodal Trefoil knot circuit with
\begin{eqnarray}
h_x({\bm k}_{\rm 3D})&=&-6 \cos  k_x \cos  k_y \cos  2k_z-0.6 \cos  k_x \cos  k_y-3 \cos^2 k_x \cos  2k_z+12 \cos  k_x \cos  2k_z-6 \cos  k_x \cos  k_z \cos  2k_z\notag\\
&&-0.6 \cos  k_x \cos  k_z-\sin^2 k_x+1.2 \cos  k_x-3 \cos^2 k_y \cos  2k_z+12 \cos  k_y \cos  2k_z-6 \cos  k_y \cos  k_z \cos  2k_z\notag\\
&&-0.6 \cos  k_y \cos  k_z+\sin^2 k_y+1.2 \cos  k_y+\cos ^3 2k_z+0.3 \cos^2 2k_z-3 \cos^2 k_z \cos  2k_z+12 \cos  k_z \cos  2k_z\notag\\
&&-12 \cos  2k_z+1.2 \cos  k_z-1.2,
\end{eqnarray}

\begin{eqnarray}
h_z({\bm k}_{\rm 3D})&=&-6 \cos  k_x \cos  k_y \cos  k_z-2 \sin  k_x \sin  k_y-3 \cos^2 k_x \cos  k_y-3 \cos  k_x \cos^2 k_y+12 \cos  k_x \cos  k_y-3 \cos^2 k_x \cos  k_z\notag\\
&&-3 \cos  k_x \cos^2 k_z+3 \cos  k_x \cos^2 2k_z+12 \cos  k_x \cos  k_z+0.6 \cos  k_x \cos  2k_z-\cos ^3 k_x+6 \cos^2 k_x-11.97 \cos  k_x\notag\\
&&-3 \cos  k_y \cos^2 k_z+3 \cos  k_y \cos^2 2k_z-3 \cos^2 k_y \cos  k_z+12 \cos  k_y \cos  k_z+0.6 \cos  k_y \cos  2k_z-\cos ^3 k_y+6 \cos^2 k_y\notag\\
&&-11.97 \cos  k_y-\cos ^3 k_z+6 \cos^2 k_z+3 \cos  k_z \cos^2 2k_z-6 \cos^2 2k_z-11.97 \cos  k_z+0.6 \cos  k_z \cos  2k_z\notag\\
&&-1.2 \cos  2k_z+7.94\notag\\
\end{eqnarray}
\pagebreak

\section{Further details on imaging Seifert surfaces through topolectrical resonance}\label{app:resonance}

\subsection{Implementation and effect of disorder}

To account for small disorder which breaks translation symmetry, Eq.~13 in the main text needs to be modified
to the following form
\begin{eqnarray}
V_a({\bm k}_{\rm 3D})&=&\sum_{\bold r}e^{-i\bold r\cdot {\bm k}_{\rm 3D}}V_{a,\bold r}=\sum_{b,{\bm k}'_{\rm 3D}}J_{ab}^{-1}({\bm k}_{\rm 3D},{\bm k}'_{\rm 3D})I_b({\bm k}'_{\rm 3D})\notag\\
&=& \sum_{n;b,{\bm k}'_{\rm 3D}}\frac{\left[|\psi_n({\bm k}_{\rm 3D},{\bm k}'_{\rm 3D})\rangle\langle\psi_n({\bm k}_{\rm 3D},{\bm k}'_{\rm 3D})|\right]_{ab}}{j_n({\bm k}_{\rm 3D},{\bm k}'_{\rm 3D})}I_b({\bm k}'_{\rm 3D}).\notag\\
\label{IJV1}
\end{eqnarray}
This recovers Eq.~13 in the main text if we assume $j_n({\bm k}_{\rm 3D},{\bm k}'_{\rm 3D})\approx j_n({\bm k}_{\rm 3D})\delta_{{\bm k}_{\rm 3D},{\bm k}'_{\rm 3D}}$, where $j_n({\bm k}_{\rm 3D})$ represents the 3D bandstructure of the clean circuit Laplacian.
Clearly then, the effect of disorder can be modeled by changing $\delta_{{\bm k}_{\rm 3D},{\bm k}'_{\rm 3D}}$ to $\delta_{{\bm k}_{\rm 3D},{\bm k}_{\rm 3D}+\Delta {\bm k}}$ in the expression of $j_n({\bm k}_{\rm 3D},{\bm k}'_{\rm 3D})$,
with $\Delta {\bm k}$ representing a density-wave disorder to the Laplacian and hence the non-diagonal expression $J({\bm k}_{\rm 3D},{\bm k}'_{\rm 3D})$ in the main text. $J({\bm k}_{\rm 3D},{\bm k}'_{\rm 3D})$ couples different wavevectors of the potential and current distributions and smudge the nodal structures. Additionally, it will also attenuate peaks in $j_n({\bm k}_{\rm 3D},{\bm k}'_{\rm 3D})$, leading to lower resonances.

\subsection{Seifert surface simulation results}
Here, to facilitate experiments in the near future we present more detailed results showcasing the behavior of the momentum-space impedance peaks (resonances) across different layers parallel to the 3D boundary layer. We denote by $Z_{ab}$ the impedance across layers $a$ and $b$, such that $Z_{11}$ is the impedance within layer $1$, $Z_{14}=Z_{41}$ is the impedance between layers $1$ and $4$, etc. Overall, it is evident that $Z_{11}$ most accurately reproduces the analytically corroborated Seifert surface, verifying that the latter is indeed a topological boundary phenomenon. Resonance peaks deeper in the bulk, i.e. $Z_{44}$ converges to the bulk NLs, also verifying that the resonances indeed originate from zero modes of the Laplacian.

For the 2-link circuit as considered in the main text, we present simulation results using a lattice with $64\times 64\times 64$ unit cells.   Results are shown
in Fig.~\ref{meaxxaa}, Fig.~\ref{meaxxab}, and Fig.~\ref{meaxyaa}.

To better understand the universal aspects of our Seifert surface imaging approach, we also simulated similar resonance measurements with a Trefoil knot nodal system in a lattice with $64\times 64\times 64$ unit cells. Results are detailed in Fig.~\ref{trefoilxxaa} and Fig.~\ref{trefoilxyaa}. Due to inversion symmetry about the $k_z=0$ plane, we have only plotted the $k_z>0$ region.   

\subsection{Chemical potential shift from grounding}

In Fig.~\ref{deaxxaa} and  Fig.~\ref{deaxyaa}, we show the results for {the same 2-link circuit} as in the main text, but with grounding capacitors that introduce an effective chemical potential~\cite{circuit3_SI} of $\mu=1$ that shifts the zero modes to other Laplacian eigenvalues. The Seifert surface, though distorted, still survives for surface layers $(a,b)=(1,1)$. However, the bulk nodal lines are now thickened into ``Fermi surface tubes''.

\begin{figure*}[h]
\includegraphics[width=1 \linewidth]{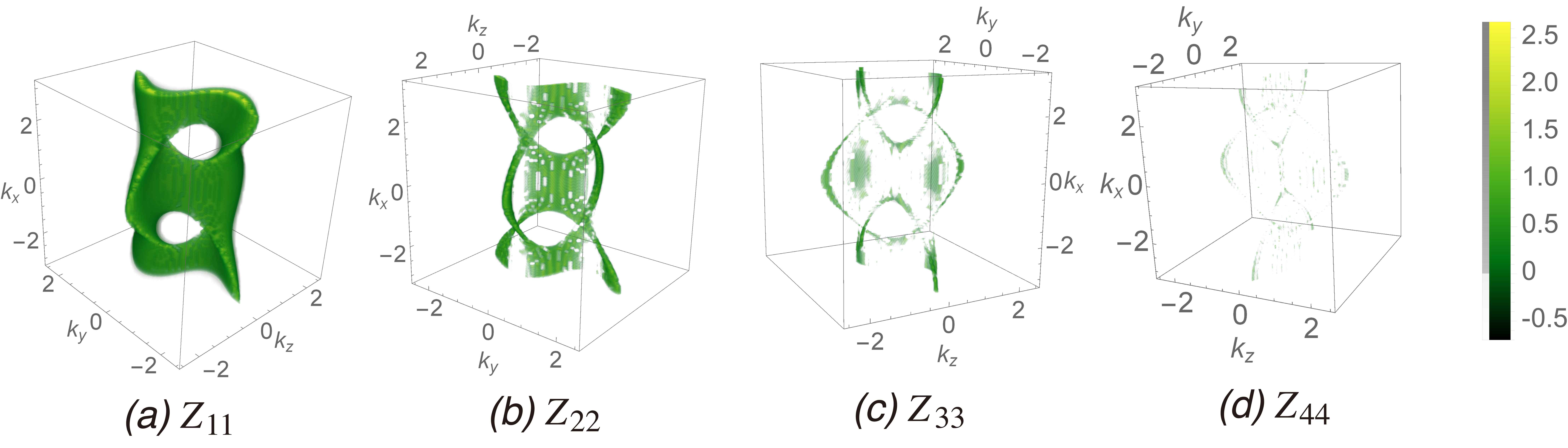}
\caption{Momentum-resolved topolectrical impedance peaks across the same sublattice and unit cell for the 2-link circuit, showing an evolution from the Seifert surface to its boundary 2-link from the 4D surface ($Z_{11}$) to the 4D bulk ($Z_{44}$). }
\label{meaxxaa}
\end{figure*}

\begin{figure*}
\includegraphics[width=1 \linewidth]{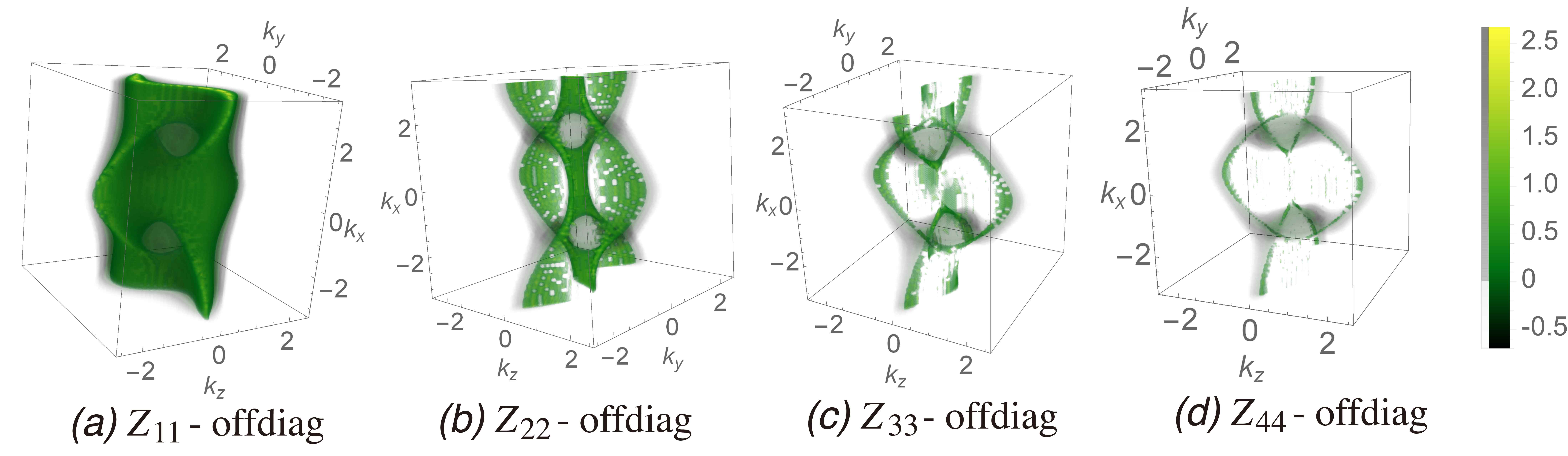}
\caption{Similar measurements as in Fig.~\ref{meaxxaa} for the 2-link circuit, but across dissimilar sublattices within the same unit cell. In some instances, the Seifert surface or bulk NL feature more prominently than in Fig.~\ref{meaxxaa}, suggesting that the off-diagonal degrees of freedom also contain crucial topological information. }
\label{meaxxab}
\end{figure*}

\begin{figure*}
\includegraphics[width=1 \linewidth]{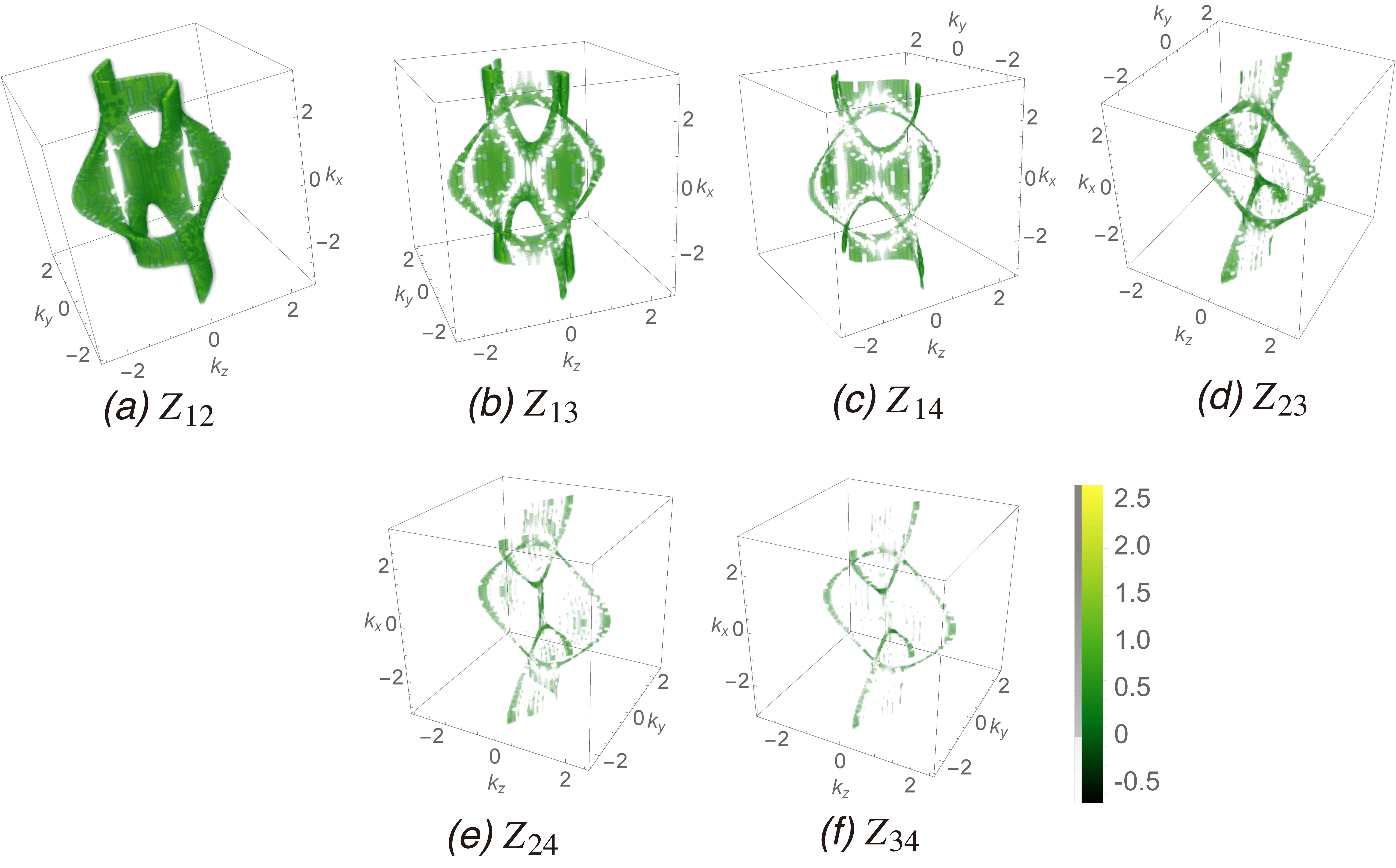}
\caption{Similar measurements as in Fig.~\ref{meaxxaa} for the 2-link circuit, but across the same sublattices of dissimilar unit cells. We observe certain ``hybrid" features between the Seifert surfaces and their bulk NLs, whose strengths indicate the extents of coupling between the bulk and the boundary modes. }
\label{meaxyaa}
\end{figure*}

\begin{figure*}
\includegraphics[width=1 \linewidth]{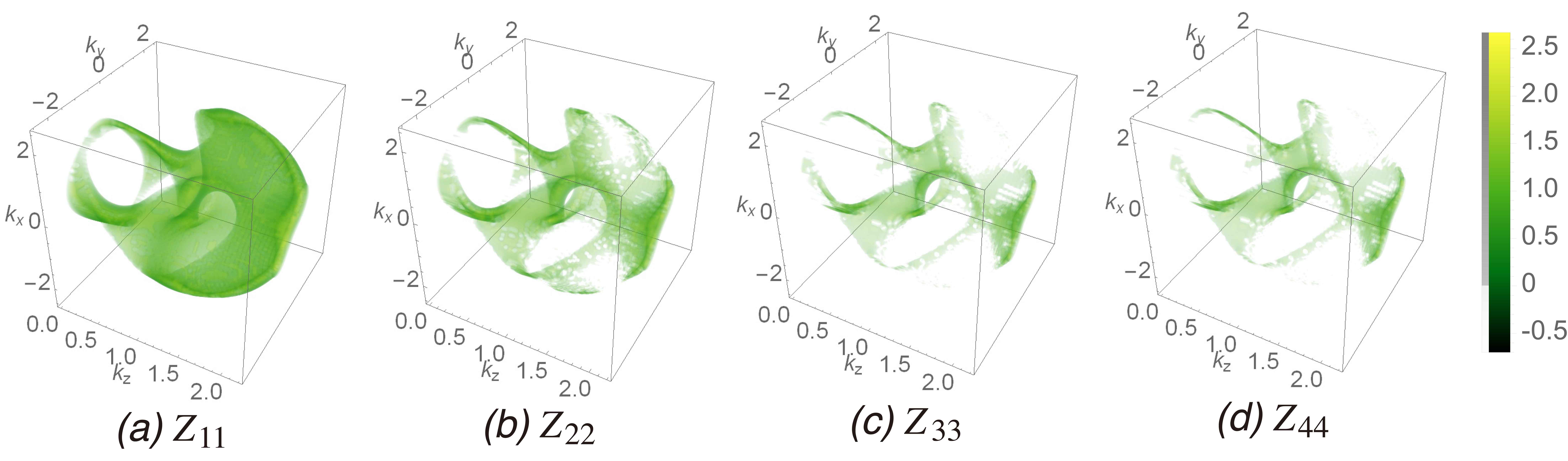}
\caption{Similar measurements as in Fig.~\ref{meaxxaa}, but for the Trefoil circuit. Like for the 2-link, we also observe the Seifert surface morphing into the bulk NLs as we move into the bulk. }
\label{trefoilxxaa}
\end{figure*}

\begin{figure*}
\includegraphics[width=1 \linewidth]{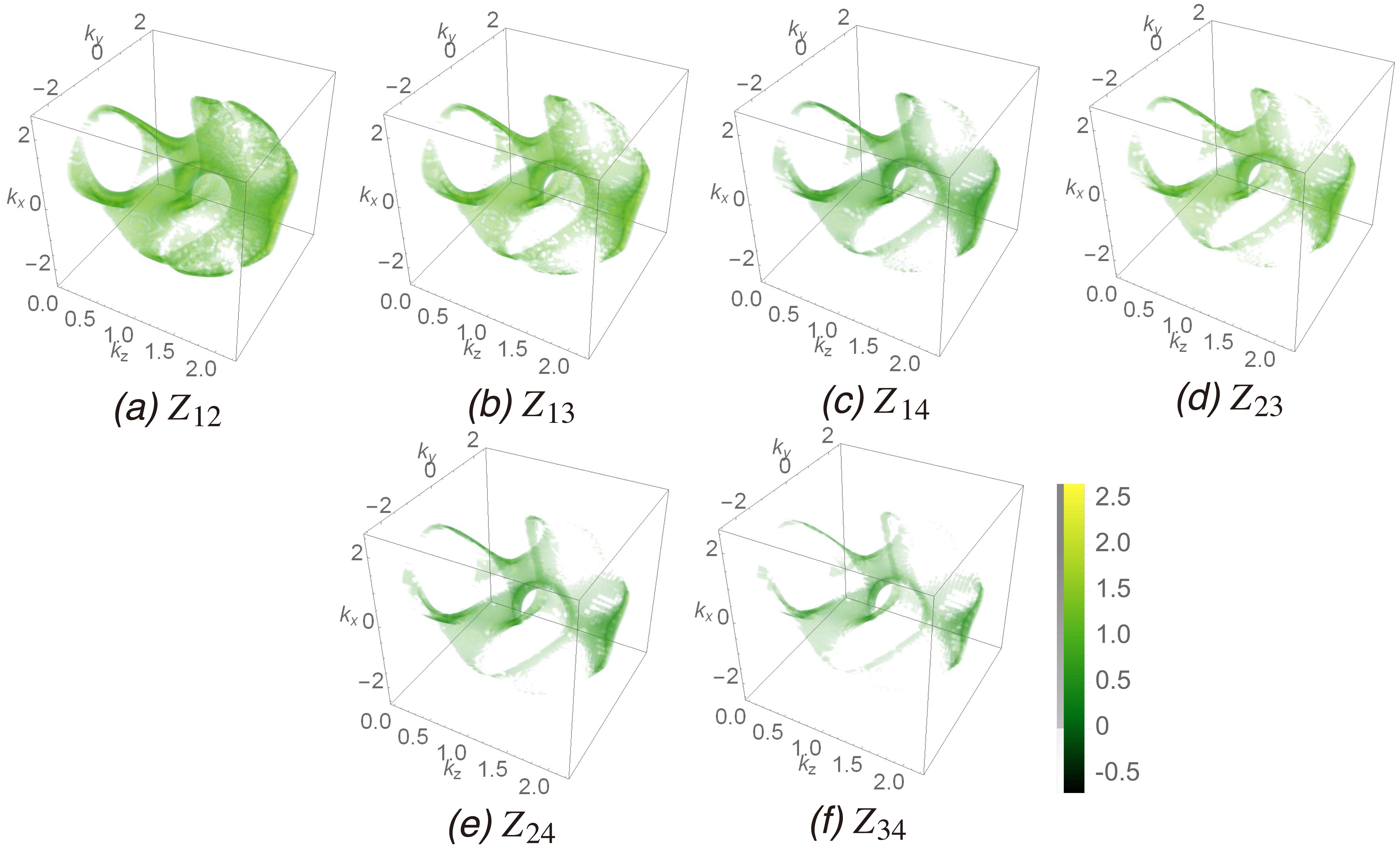}
\caption{Similar measurements as in Fig.~\ref{meaxyaa}, but for the Trefoil circuit. To eradicate remnants of the Seifert surface in the bulk, one either has to venture deeper into the bulk, or use a cleaner (less disordered) circuit.}
\label{trefoilxyaa}
\end{figure*}

\begin{figure*}
\includegraphics[width=1 \linewidth]{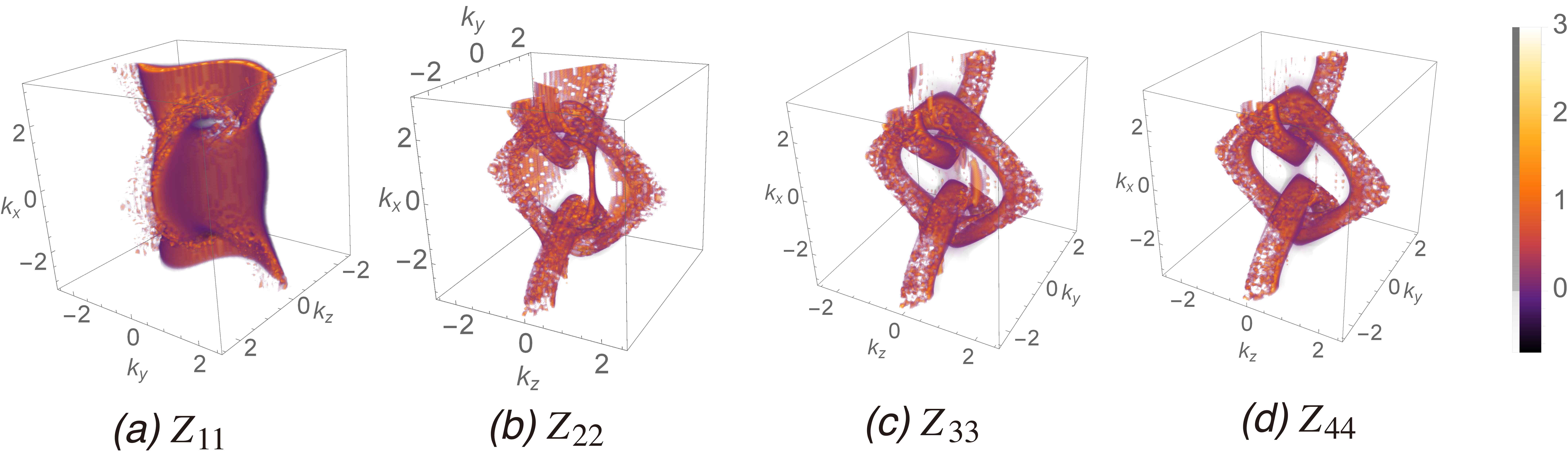}
\caption{Diagonal resonances across the same unit cell for the 2-link, but at a chemical potential $\mu=1$ introduced via grounding capacitors. While the Seifert surface in a) has become distorted, it is still bounded by the thickened NLs, in exact agreement with the schematic illustration in Fig. 4(c) in the main text.}
\label{deaxxaa}
\end{figure*}

\begin{figure*}
\includegraphics[width=1 \linewidth]{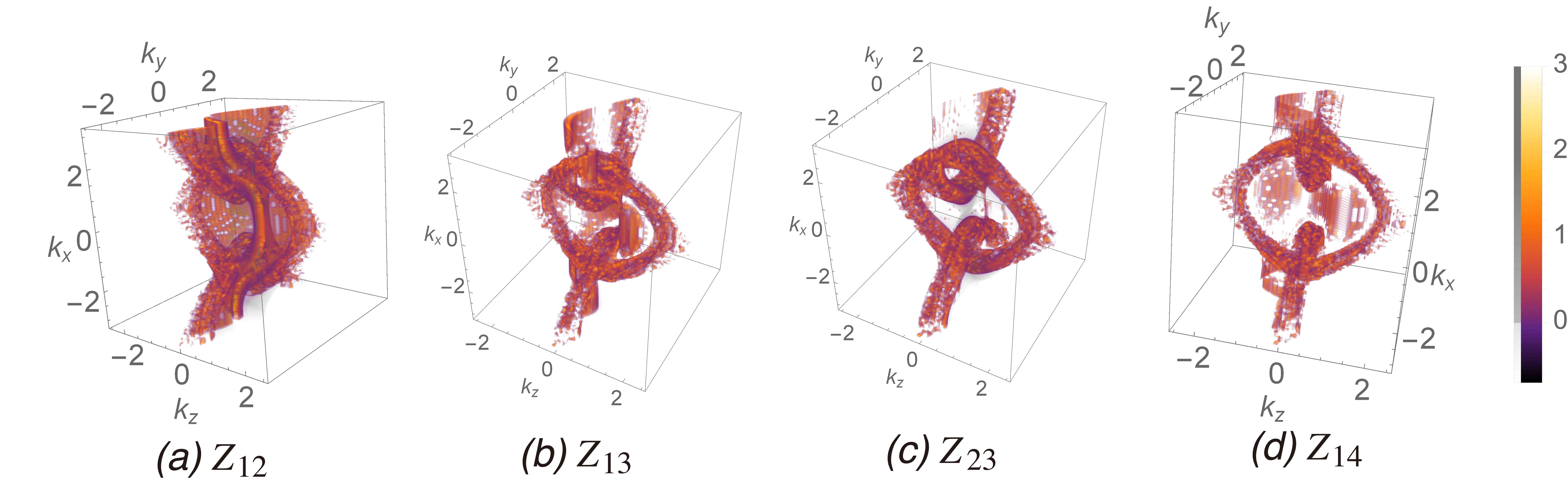}
\caption{Diagonal resonances across the same unit cell for the 2-link at $\mu=1$, with the noisy texture attributed to poor boundary localization away from the gap closure point, which is picked up when $\mu=0$.}
\label{deaxyaa}
\end{figure*}


\begin{thebibliography}{99}
\bibitem{footnote1}
An alternative Seifert surface can be defined by the loci of equal pseudospin direction~\cite{Seifert_light}, although it cannot be directly measured from the band structure.
\end{thebibliography}
\end{document}